\documentclass{egpubl}
\usepackage{egsgp2020}

 \SpecialIssuePaper         

\usepackage[T1]{fontenc}
\usepackage{dfadobe}  
\usepackage{pgfplots}
\usepackage{amsmath}
\usepackage{algorithmicx}
\usepackage{algorithm}
\usepackage{algpseudocode}
\usepackage{wrapfig}
\usepackage{floatflt}
\usepackage{units}
\usepackage[normalem]{ulem} 

\usepackage{amsfonts}
\newcommand{\R}{{\mathbb{R}}}

\newcommand{\IP}{\mathrm{IP}}
\newcommand{\length}{\mathrm{length}}
\newcommand{\area}{\mathrm{area}}
\newcommand{\dist}{\mathrm{dist}}
\newcommand{\MA}{\mathrm{MA}}
\newcommand{\MAG}{\mathrm{MAG}}

\newcommand{\inr}{\mathrm{inr}}
\newcommand{\inx}{\mathrm{inx}}
\newcommand{\inB}{\mathrm{inB}}

\newcommand{\add}[1]{{#1}}
\newcommand{\deletesec}[1]{{}}
\DeclareRobustCommand{\delete}[1]{{}}

\usepackage{amsthm}
\theoremstyle{definition}
\newtheorem{definition}{Definition}[section]
\newtheorem{proposition}{Proposition}

\usepackage{cite}  
\BibtexOrBiblatex
\electronicVersion
\PrintedOrElectronic
\ifpdf 
\usepackage[]{graphicx} \pdfcompresslevel=9
\else \usepackage[dvips]{graphicx} \fi

\usepackage{egweblnk} 
\usepackage{mathtools}

\title{Medial Axis Isoperimetric Profiles}

\author[Zhang, DeFord, \& Solomon]
{\parbox{\textwidth}{\centering P.\ Zhang,$^{1}$\orcid{0000-0003-4136-1315} D. DeFord,$^{1, 2}$ 
        and J. Solomon$^{1}$\orcid{0000-0002-7701-7586}
        }
        \\
{\parbox{\textwidth}{\centering $^1$Massachusetts Institute of Technology, USA
       }
}
\\
{\parbox{\textwidth}{\centering $^2$Washington State University, USA
       }
}
}

\begin{document}


\maketitle
\begin{abstract}

Recently proposed as a stable means of evaluating geometric compactness, the \emph{isoperimetric profile} of a planar domain measures the minimum perimeter needed to inscribe a shape with prescribed area varying from 0 to the area of the domain.  While this profile has proven valuable for evaluating properties of geographic partitions, existing algorithms for its computation rely on aggressive approximations and are still computationally expensive. In this paper, we propose a practical means of approximating the isoperimetric profile and show that for domains satisfying a ``\emph{thick neck}'' condition, our approximation is exact. For more general domains, we show that our bound is still exact within a conservative regime and is otherwise an upper bound. Our method is based on a traversal of the medial axis which produces efficient and robust results. We compare our technique with the state-of-the-art approximation to the isoperimetric profile on a variety of domains and show significantly tighter bounds than were previously achievable.


\end{abstract}  

\vspace{-5px}
\section{Introduction}

The \emph{isoperimetric problem} can be traced back to \emph{Dido's problem}, in which Dido---the founder of Carthage---wished to maximize her territory while bounding it with a limited supply of strips of bull's hide \cite{Aeneid}. This problem is solved by invoking the \emph{isoperimetric inequality}, which states for any shape of perimeter $P$ and area $A$, $P^2\geq 4\pi A$; as a result, her encapsulated territory is maximized by a circle. 

The isoperimetric inequality motivates the definition of the \emph{isoperimetric ratio} $\frac{4\pi A}{P^2}$, which quantifies compactness of a shape. In the study of geography and political redistricting, the word \emph{compactness} is used informally to distinguish shapes that are not too irregular or distorted, rather than in the formal mathematical sense. For example, a perfect circle is the most compact shape achievable, with isoperimetric ratio $1$, while a shape with fractal-esque boundary will have a ratio close to $0$. Beyond serving as a basic tool in analytic geometry, the isoperimetric ratio---known as the \emph{Polsby--Popper score} in political science \cite{Polsby1991TheTC}---has found application in comparing geographic partitions used in political redistricting.

\begin{figure}
    \centering
    \input{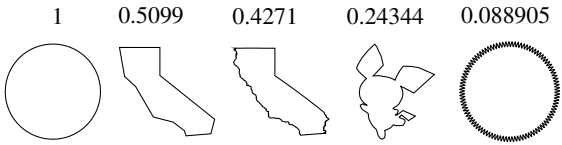}
    \label{fig:ppscore}
    \caption{From left to right: circle, low-res-California, high-res-California, silhouette, perturbed-circle. Their isoperimetric ratios are denoted above them. This indicates that a large range of shapes can lie between essentially two circles on the isoperimetric ratio scale. }
\end{figure}

Applications of the isoperimetric ratio to geographic domains suffer from the ``coastline paradox.''  In particular, since geographic boundaries can have fractal shapes, their lengths are not well-defined and can change at different length scales. 
The isoperimetric ratio is unstable to boundary perturbation and will change its compactness score depending on the resolution of the input shape (see Figure \ref{fig:ppscore}). 
This instability can have a severe impact on applications
to political redistricting \add{where a low compactness score is frequently presented as evidence of improper intent in the line drawing process} \cite{bar2019gerrymandering,Barnes_Solomon}. \add{Additionally, when alternative plans are considered in court, scores like Polsby--Popper are used to evaluate whether a proposed plan would be a permissible remedy. 
While straightforward geometric measures are convenient for making direct pairwise comparisons in court, their simplicity and inherent instability means that they are exploitable and insufficient to distinguish between multiple potential types of undesirable shapes. }

Recent works have introduced variations of the classical isoperimetric problem to address the shortcomings above. In particular, the \emph{isoperimetric profile} (IP) modifies the original problem by producing a function rather than a single value. 
Intuitively, the isoperimetric profile augments the isoperimetric ratio by outputting a geometric compactness score for all length scales rather than at just one. More formally, the profile of a domain 
$\Omega \subset \R^2$ is defined via the following optimization: 
\begin{equation}\label{eq:IP}
\IP_{\Omega}(t) := 
\min \left\{ \length(\partial F) : F \subseteq \Omega 
\And
\area(F) = t
\right\},\tag{IP}
\end{equation}
where $F$ is an inscribed shape within $\Omega$, $\length(\partial F)$ is the perimeter of $F$, $\area(F)$ is the area of $F$, and $t\in[0,\area(\Omega)]$ is the multi-scale resolution parameter of the profile. The profile for a given value of $t$ is the smallest perimeter a shape $F$ inscribed in $\Omega$ can have while also maintaining area $t$. There is no restriction that $F$ must be connected. Following the notation in \cite{giorgio2019}, we will refer to the $F$ that solves this optimization as $E_{\Omega}(t)$. 

\begin{figure}
    \centering
    \includegraphics[width=1\columnwidth]{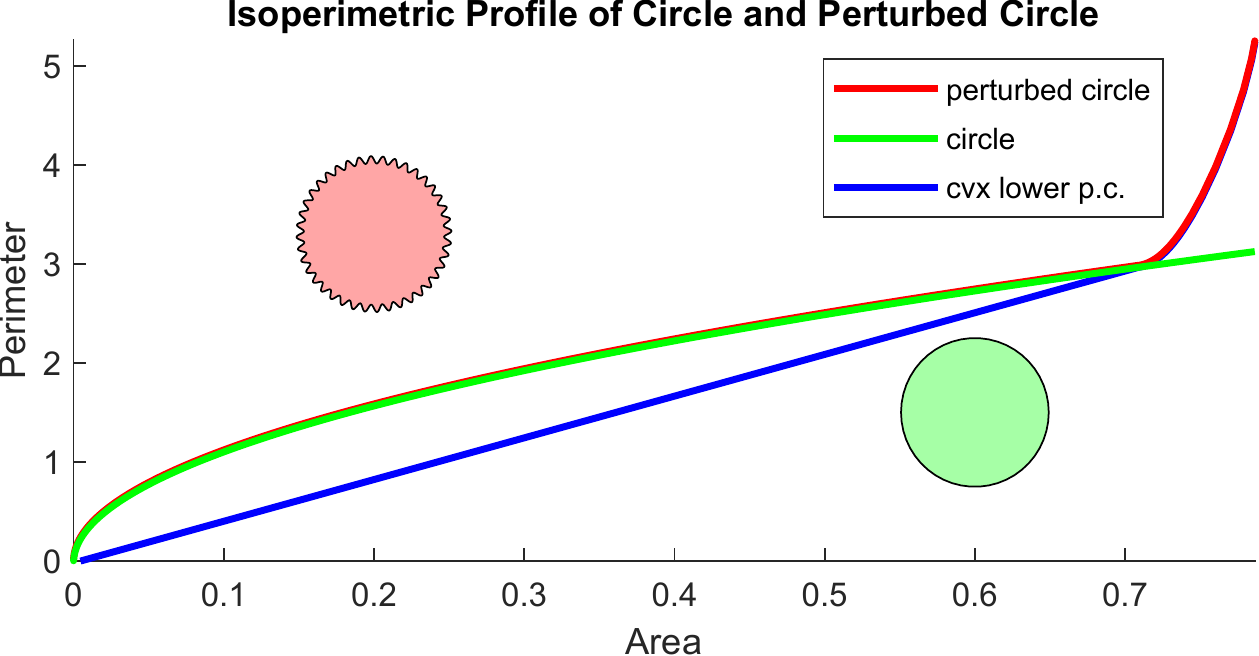}
    \caption{Isoperimetric profile plotted for two similar shapes of the same area: (green) the circle, and (red) a perturbed circle. The profile shows that these two shapes are identical at large length scales, but differ significantly on smaller length scales. The blue curve shows what \cite{DeFord2019TotalVI} computes: the convex lower envelope of the profile of the perturbed circle. Note the circles are not plotted on the same scale as the profile.}
    \label{fig:simpleprofiles}
\end{figure}

While the isoperimetric ratio is a scalar measurement of compactness, the isoperimetric profile provides an entire plot that measures the \emph{multiscale compactness} of a domain. As such, it is more suited to noisy data such as geographic domains that often exhibit fractal boundaries on coastlines and rocky territories. The notion of multiscale compactness is demonstrated in Figure \ref{fig:simpleprofiles} for two near-circular shapes that manifest extreme isoperimetric ratio values. While the ratio strongly distinguishes these two shapes, the profile reveals that they are in fact similar at coarse length scales and only differ at fine length scales. \add{Furthermore, the isoperimetric profile is invariant to translation and rotation of the domain. When the areas of different domains are normalized, their isoperimetric profiles can be used as multiscale shape descriptors for measuring similarity and clustering.}

Despite the theoretical advantages of the isoperimetric profile, \add{the lack of an}\delete{no} efficient algorithm \delete{is known} to compute it \add{heavily deters its usage in practice}. A recent method takes a convex relaxation of Eq.~\eqref{eq:IP} using \emph{total variation} (TV) on an Eulerian grid to compute the convex lower envelope of $\IP_{\Omega}[t]$ \cite{DeFord2019TotalVI}. This method suffers from several drawbacks: the lower bound is not tight on all valid values of $t$ for any domain; the discretization converges poorly under mesh refinement; and computation of $\IP_{\Omega}[t_1]$ does not leverage results from previously computed values $\IP_{\Omega}[t < t_1]$, although we note that this final issue could be possibly addressed by using a solution for smaller $t$ as the starting point for the next round of optimization. 

{}

{}

The drawbacks in \cite{DeFord2019TotalVI} limit its applicability to accurately obtaining the isoperimetric profile. To rectify this, we present a novel algorithm that computes upper bounds on the isoperimetric profile that are provably tight for a restricted class of domains. Outside this class, we still attain tight bounds on a subset of scales $t$. Our algorithm 
exploits restrictions on the mean curvature of $\partial E_{\Omega}(t)$ \cite{gonzalez1980regularity}
as well as recent results on values of $E_{\Omega}[t]$ for special domains \cite{giorgio2019}. This leads us to the construction of our bound via the medial axis of $\Omega$. We use our algorithm to bound isoperimetric profiles for a large variety of domains, including districts, states and countries at various resolutions. We show---on domains for which the isoperimetric profile is known---that our bound is much tighter than bounds computed by previous state-of-the-art.

Our main technical contributions are as follows:
\begin{itemize}
    \item We generalize previous approaches to understanding the isoperimetric profile by constructing it through the medial axis.
    \item We design a novel algorithm for efficiently computing upper bounds to the isoperimetric profile.
    \item We prove that our bound is tight for a restricted set of domains.
    \item We prove that on general domains $\Omega$ our bound is tight for a range of $t$ values derived from properties of $\Omega$.
    \item We provide experiments demonstrating that our bounds on the isoperimetric profile are tight enough to be used in practice.
\end{itemize}{}

\vspace{-10px}
\section{Related Work}
While the original isoperimetric problem has been solved for centuries, many variants remain open and actively studied. For a summary of results on the subject in 2D, on Riemannian 3-manifolds, and on measures, see \cite{antonio2005}. 

A key application of the isoperimetric ratio is as a way to measure geometric compactness to detect gerrymandering \cite{Polsby1991TheTC}. In this context, it is known as the  \emph{Polsby-Popper} score. In some U.S.\ states, it is even a legally required statistic to be reported for any proposed districting plan \cite{MinessotaOrder}. Despite this, \cite{kaufmanLeg, duchin2018discrete,bar2019gerrymandering,Barnes_Solomon} show that this score is not a reliable measure of compactness. \cite{duchin2018discrete} suggest a discrete Polsby-Popper score on graphs that alleviates problems with map projection at the cost of geometric sensitivity. \cite{Ehrenburg1982} suggest a variation of the isoperimetric ratio that computes the ratio of the area of the full domain to the area of its maximum inscribed circle. Ultimately, this measure ignores high resolution features of the domain. 
\add{Note that the Ehrenburg test can be extracted from the front of the isoperimetric profile, while the Polsby-Popper score can be extracted from its end with interpolation in between.}
\delete{Both The Ehrenburg test and the Polsby-Popper score can be extracted from single points on the isoperimetric profile}.

Many variations of the isoperimetric problem differ from the isoperimetric profile we consider in that they use the \emph{relative perimeter} of the inscribed domain rather than \emph{full perimeter}. While the relative perimeter formulations do not count shared boundary between the domain and its inscribed subdomain \cite{antonio2005}, the full perimeter does.
\cite{Stredulinsky1998} provide conditions under which a perimeter-minimizing subdomain of a convex set is also convex.
\cite{giorgio2019} study the maximizers of a curvature functional with close ties to the isoperimetric profile when restricted to domains with no necks. Their results provide a method for computing the isoperimetric profile when restricted to no neck domains.
\cite{DeFord2019TotalVI} propose the first algorithmic approach to the isoperimetric profile on general domains by using TV as a generalized measurement of perimeter of a domain. This relaxation of the problem yields the convex lower envelope of the isoperimetric profile. 
It remains unknown if computation of the exact isoperimetric profile for 2D polygons is in the polynomial complexity class. Additional open problems for the 2D profile are summarized in \cite{croft2012unsolved}. 

\vspace{-10px}
\subsection{Properties of solutions to \eqref{eq:IP}}
\label{sec:propsolIP}
While there is no known efficient algorithm to compute the full isoperimetric profile, much is known about $\partial E_{\Omega}(t)$. In particular, for any fixed value of $t$, the \add{\emph{free boundary}} \delete{boundary} of $E_{\Omega}(t)$ \delete{away from $\partial \Omega$}, $\partial E_\Omega(t)\setminus \partial \Omega$ must have constant and nonsingular mean curvature \cite{gonzalez1980regularity}. In addition, $\partial E_{\Omega}(t)$ and $\partial \Omega$ must meet tangentially, and any free boundary of $E_{\Omega}(t)$ must be a circular arc \cite{BESICOVITCH1949}. For a small subset of values of $t$, $E_{\Omega}(t)$ and $\IP_{\Omega}(t)$ are known in closed form \cite{giorgio2019}. We will expand on this in \S\ref{sec:morphopenprelim} after establishing more notation and geometric tools.

\vspace{-10px}
\section{Preliminaries}
\subsection{Domains}
Literature on the isoperimetric profile has typically focused on suitably regular \emph{Jordan domains}. 
\begin{definition}[Jordan Curve \cite{sulovský2012depth}]
A \emph{Jordan curve} or a \emph{simple closed curve} in the plane is the image of an injective continuous map $\phi:\mathbb{S}^1\rightarrow \R^2$.
\end{definition}{}
\begin{definition}[Jordan Domain \cite{sulovský2012depth}]
A \emph{Jordan domain} is the interior of a Jordan curve.
\end{definition}{}
Beyond restricting attention to Jordan domains, there is typically an additional regularity constraint that  $\area(\partial\Omega)=0$, where $\area(\cdot)$ denotes the Lebesgue 2-dimensional measure. 
Domains satisfying this condition include any finite polygon and Jordan domains with smooth boundary. For contrast, \emph{Osgood curves} are Jordan curves that can have nonzero area \cite{Knopp1917}, but for our purposes these will not be considered. The set of suitably regular Jordan domains will be denoted $\mathbb{J}$.

\vspace{-5px}
\subsection{Necks}
We will further refine these domains $\Omega$ by the characteristics of their \emph{necks}. 
\begin{definition}[Neck, \cite{CheegerLeonardi2017}]
\label{def:neck}
A domain $\Omega\subset\R^2$ has a \emph{neck} of radius 
$\rho$ if 
there exists a pair of balls 
$B_{\rho}(x_0),\;B_{\rho}(x_1) \subseteq \Omega$ such that there does not exist a 
continuous curve 
$\gamma:[0,1]\rightarrow\Omega$ satisfying $\gamma(0)=x_0$, $\gamma(1)=x_1$, and $B_{\rho}(\gamma)\subseteq\Omega$.
\end{definition}
Let $r_n$ denote the smallest value of $\rho$ for which $\Omega$ has a neck. This definition is built intuitively on the idea of a bottleneck where objects of a certain size are unable to pass through. A domain that does not satisfy Definition \ref{def:neck} for any $\rho$ has \emph{no neck}, i.e., $r_n=\infty$. For example, \emph{star} domains have no necks. In contrast, it is clear that the shape in Figure \ref{fig:mopen} has $r_n < \infty$.  This definition does not have a concept of multiple necks nor does it take into account the locations of necks. We will provide a new definition in \S \ref{sec:medaxisnecks} that generalizes necks to incorporate these concepts. 

\vspace{-5px}
\subsection{Morphological Operations}
\label{sec:morphopenprelim}

\begin{figure}
    \centering
    \hfill 
    \includegraphics[width=.32\columnwidth]{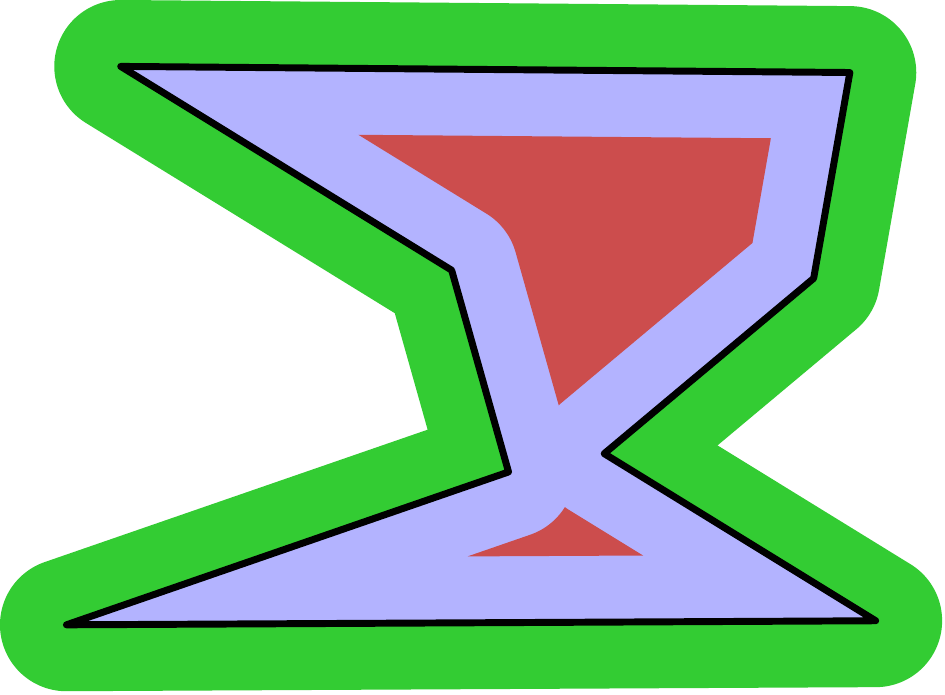}
    \hfill
    \includegraphics[width=.32\columnwidth]{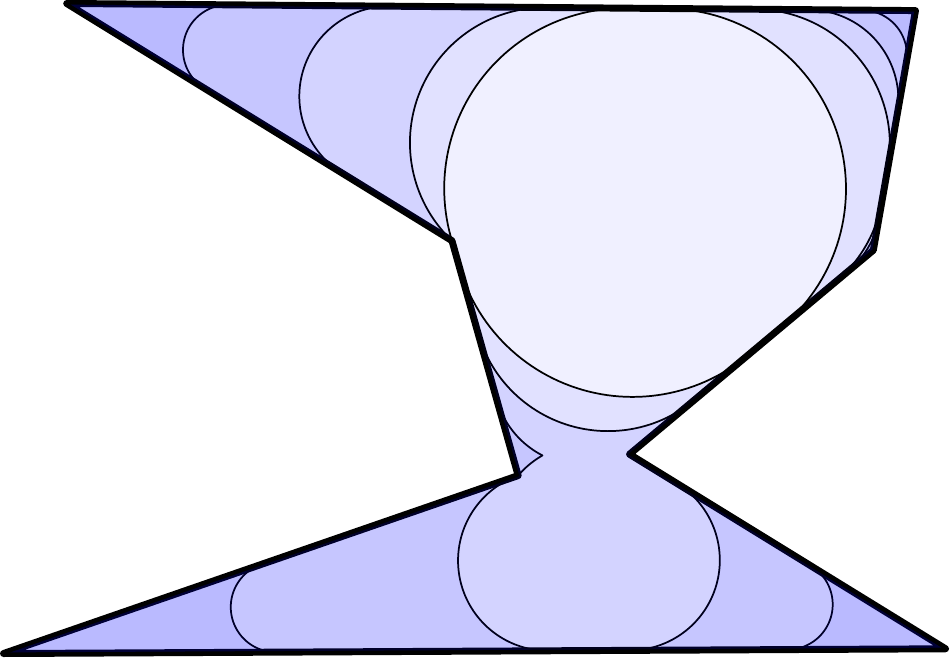}
    \hfill
    \includegraphics[width=.32\columnwidth]{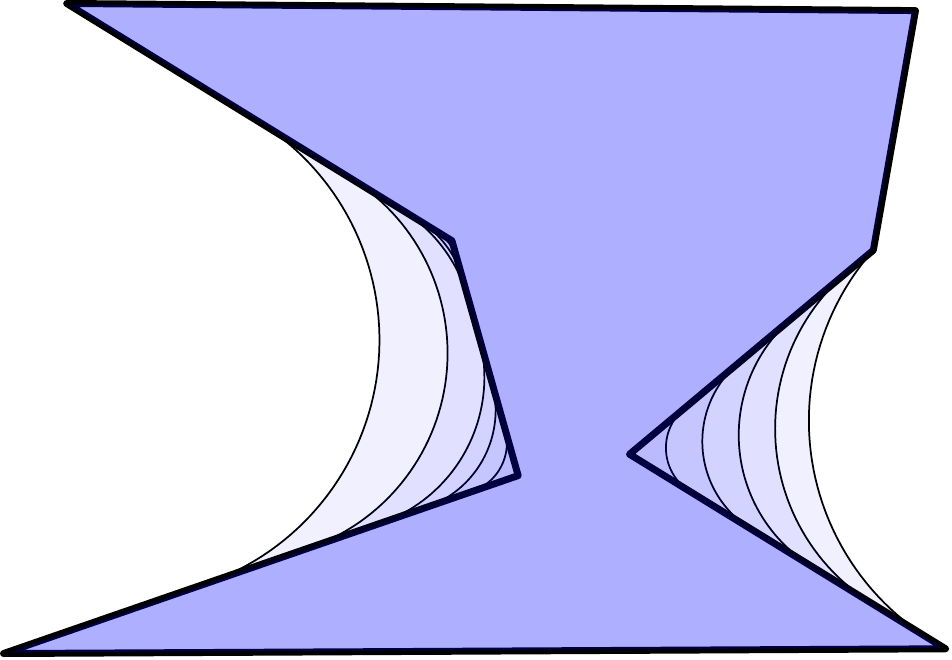}
    \hfill
    \caption{(Left) The thickly outlined blue shape is the starting domain. The innermost red domain is the erosion of the starting domain by $B_r$. The outermost green shape is the dilation of the starting domain by $B_r$.  (Middle) Morphological opening of the starting domain by balls of changing radius. Darker blue corresponds to \delete{larger}\add{smaller} radius. (Right) Morphological closing of the starting domain by balls of changing radius. \delete{Lighter}\add{Darker} blue corresponds to \delete{larger}\add{smaller} radius.}
    \label{fig:mopen}
\end{figure}

Natural variations of a shape are given by morphological operations. Furthermore $E_{\Omega}(t)$ for limited values of $t$ can be characterized by the morphological \emph{opening} of $\Omega$. We will also use morphological \emph{closing} in \S \ref{sec:empanl} to clean up our results.  For these reasons, we provide their definitions; see \cite{Serra} for more discussion.

Let $B_r(x)$ be a ball of radius $r$ centered at $x\in\R^2$ and $B_r \coloneqq B_r(0)$.
Given a starting domain $\Omega\subset\R^2$, its \emph{erosion} by $B_r$ is 
\begin{equation}
    \Omega \ominus B_r = \left\{ x\in\Omega: B_r(x) \subseteq \Omega \right\}.
\end{equation}{}
Its \emph{dilation} by $B_r$ is
\begin{equation}
    \Omega \oplus B_r = \left\{ x\in\R^2: \Omega\cap B_r(x) \neq \emptyset \right\}.
\end{equation}{}
Its opening by $B_r$ is 
\begin{equation}
    \Omega \circ B_r = (\Omega \ominus B_r) \oplus B_r.
\end{equation}{}
Finally, its closing by $B_r$ is 
\begin{equation}
    \Omega \bullet B_r = (\Omega \oplus B_r) \ominus B_r.
\end{equation}{}
These operations and their results are depicted in Figure \ref{fig:mopen}. Intuitively, the opening can be thought of as a blurring of the convex corners of a shape. The closing can be thought of as a way of filling in small holes or smoothing out concave corners of a shape.

With this notation, we recall a property of the isoperimetric profile, namely that
$E_{\Omega}(t) = \Omega \circ B_r$ when $r < r_n$ with corresponding area $t = \area(\Omega \circ B_r)$ \cite{giorgio2019}. 
It follows that when restricted to no neck domains, the isoperimetric profile can be computed by opening \cite[Theorem 2.3]{giorgio2019}. 
For domains with necks, however, the opening is not generally equal to $E_{\Omega}(t)$. This is exemplified in Figure \ref{fig:mopen} where the opening has singular curvature near the neck. Furthermore, we find that in practice the range $t \in [ \area(\Omega \circ B_{r_n}), \area(\Omega)]$ accounts for a very small portion of the isoperimetric profile on domains with necks.

\vspace{-5px}
\subsection{Medial Axis}
\label{sec:medaxprelim}
One of our key ideas is to connect the widely used \emph{medial axis} to the isoperimetric profile. The medial axis allows us to formulate a new definition of neck that will be helpful for our analysis.

\begin{figure}
    \centering
    \hfill
    \includegraphics[width=.45\columnwidth]{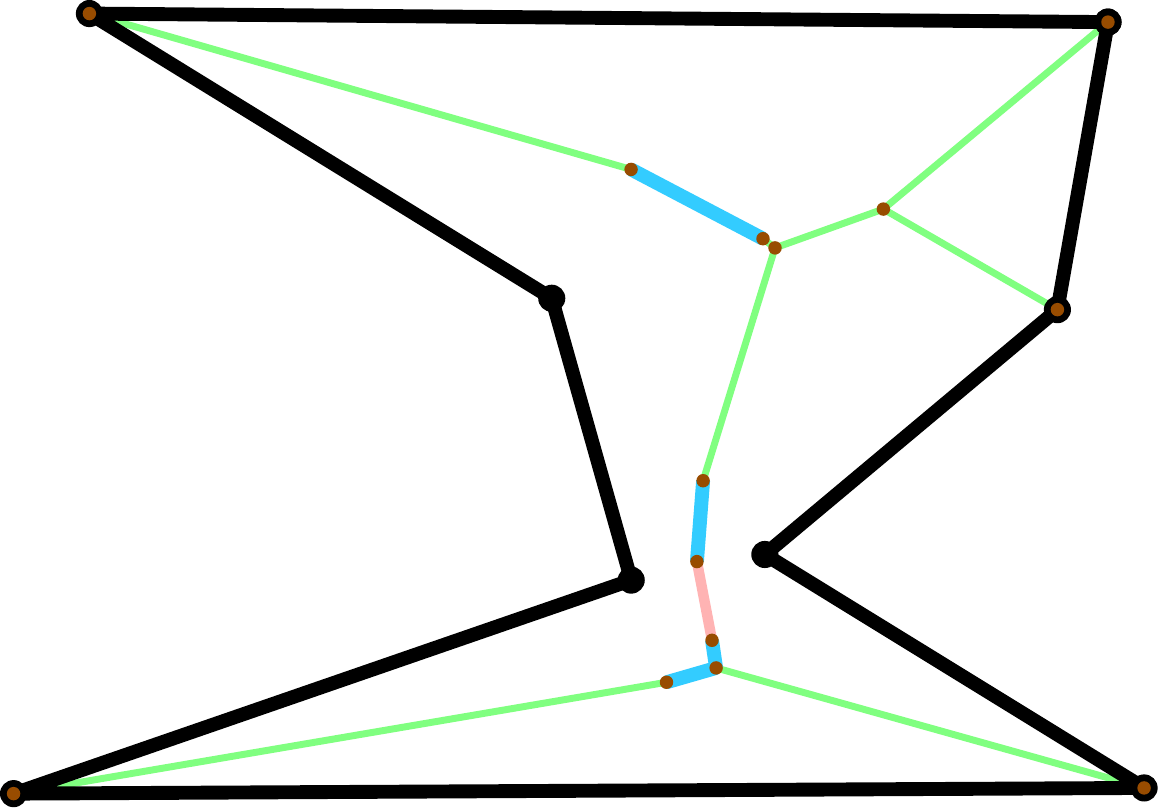}
    \hfill
    \includegraphics[width=.45\columnwidth]{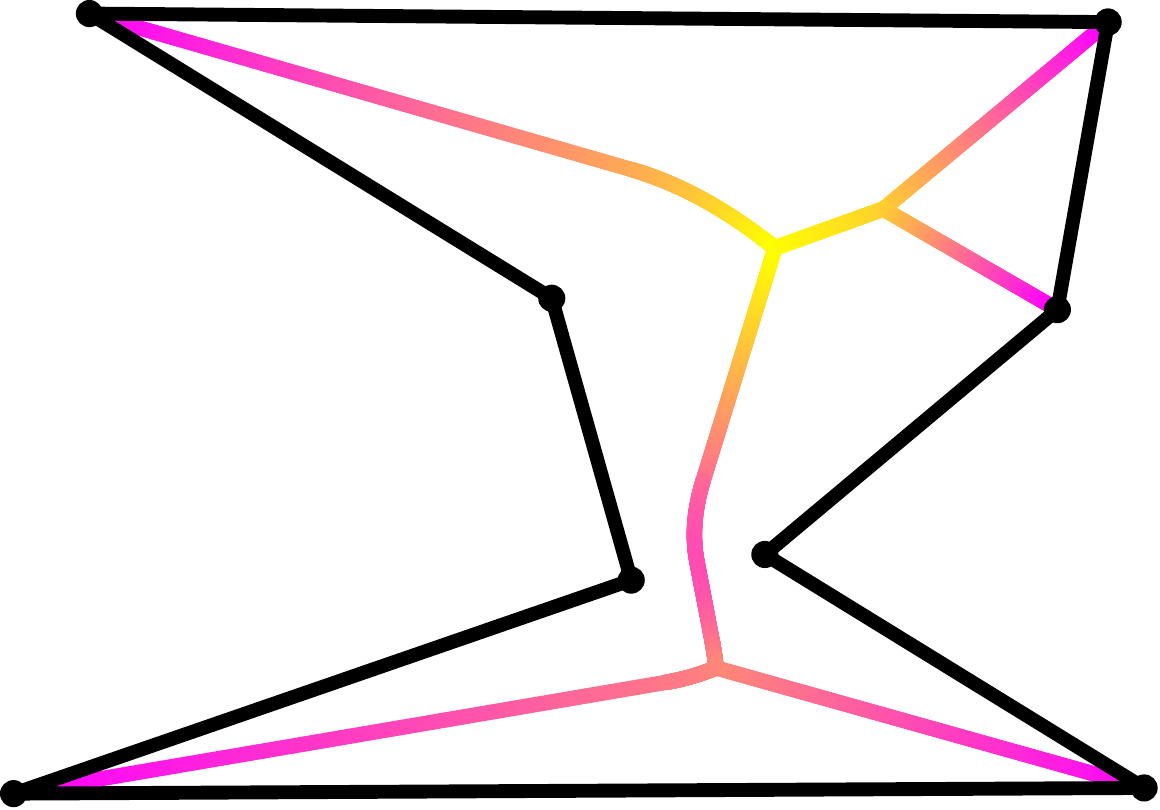}
    \hfill
    \caption{(Left) The domain $\Omega$ is outlined in black. Connectivity of medial axis edge segments are shown with colors indicating their type. Green edges are edge-edge. \delete{Red}\add{Pink} edges are vert-vert. Blue edges are edge-vert. \add{This distinction between edge segment types allows us to characterize the shape of the curve (linear or quadratic), as well as its radius function.} Brown points indicate medial axis nodes where multiple segments connect. (Right) Piecewise quadratic medial axis is shown with color indicating radius function. Light yellow indicates high radius while magenta indicates low radius.}
    \label{fig:medaxis}
\end{figure}

Intuitively, the medial axis of a 2D domain is a set of curves that characterize the skeleton of a shape. To formalize this idea, let $\Pi_{\partial \Omega}(x\in\Omega)$ be the projection 
of $x$ onto the closest boundary point.
The medial axis of $\Omega\subset\R^2$ is 
\begin{equation}
    \MA(\Omega) = 
    \left\{ 
     x\in\Omega \left\vert
    \Pi_{\partial \Omega} \mathrm{\;is\;discontinuous\;at}\;x
    \right.
    \right\}.
\end{equation}{}
\noindent Said differently, the medial axis contains all interior points where the closest boundary point is non-unique. 
The medial axis of a 2D polygon can be decomposed into the union of linear and quadratic edge segments \cite{Arinyo1999ComputingTM}. Each \emph{medial axis edge segment} is a set of points that are equidistant from two boundary objects, \emph{governors}: two boundary vertices (vert-vert), two boundary edges (edge-edge), or one boundary vertex and one boundary edge (edge-vert). These edge segments join at \emph{medial axis nodes}. The \emph{medial axis transform} is the coupling of the medial axis with its radius function $r(x\in\MA(\Omega)) = \dist(x,\Pi_{\partial \Omega}(x))$, which measures its distance to the boundary. 
The medial axis transform is unique, and it can be used to reconstruct $\Omega$ \cite{Blum:1967:ATF}. 
An example medial axis is provided in Figure \ref{fig:medaxis}.
Lastly, we denote the the largest circle inscribed in $\Omega$ by
$\inB(\Omega)$ with corresponding radius $\inr(\Omega) = \max\{r(x): x\in\MA(\Omega)\}$ and center $\inx(\Omega)$. We assume $\inB(\Omega)$ is unique, which is guaranteed for a generic domain, but not for domains with too much symmetry. In such a case, a randomized boundary perturbation will give us the needed genericity. 

{}

\vspace{-5px}
\section{Motivating the Algorithm}
We build our algorithm by first considering ways of computing the isoperimetric profile on domains with no necks, and then extending them to general domains. As mentioned in \S \ref{sec:morphopenprelim}, for domains with no necks, $E_{\Omega}(t)$ is always equal to $\Omega \circ B_r$ for some $r$. Since $\area(\Omega\circ B_r)$ decreases strictly monotonically with $r$, $r$ and $t$ are in correspondence. Thus, we can build the isoperimetric profile by measuring the area and perimeter of $\Omega \circ B_r$ for values of $r\in[0,\inr(\Omega)]$. It remains to consider different algorithmic ways of building $\Omega \circ B_r$. A key consideration of our algorithm will be how well it extends to general domains with necks.

\vspace{-5px}
\subsection{Direct computation of $\Omega \circ B_r$}
\label{sec:morphopen}
The opening of a 2D domain is simple to compute from modern geometry processing tools and generates feasible solutions to problem \eqref{eq:IP} on general domains. Unfortunately, the IP upper bound generated by measuring the perimeter of $\Omega \circ B_r$ exhibits several undesirable traits on domains with necks. Firstly, $\Omega \circ B_r$ varies discontinuously with $r$, leaving large gaps in the profile; we repair this issue in \S \ref{sec:empanl}. Second, this IP bound is not monotonic, and we know the true isoperimetric profile should always monotonically increase. Finally, $\partial (\Omega \circ B_r)$ can have singular curvature on free boundaries, while $\partial E_{\Omega}(t)$ cannot. For these reasons direct computation of $\Omega \circ B_r$ does not extend well to domains with necks. To fix these shortcomings, we propose a novel medial axis based approach.

\vspace{-5px}
\subsection{Medial Axis Limited Reconstruction}
\label{sec:medaxpartrecon}
Recall from \S\ref{sec:medaxprelim} that the medial axis transform is invertible i.e. we can reconstruct $\Omega$ from $\MA(\Omega)$ and its radius function $r(x)$. Unsurprisingly, performing the reconstruction on a subset of $\MA(\Omega)$ will produce a subset of $\Omega$. We will refer to these as \emph{limited reconstructions}. 
\begin{definition}[Limited Reconstruction]
Let $g\subseteq\MA(\Omega)$ be a subset of the medial axis.
The medial axis \emph{limited reconstruction} of $\Omega$ by $g$ is 
$\Omega_g = \bigcup_{x\in g} B_{r(x)}(x)$.
\end{definition}
When $g=\MA(\Omega)$, $\Omega_g=\Omega$. It turns out that for a particular choice of $g$, we can relate the morphological opening of a domain to its limited reconstruction.
\begin{proposition}
\label{prop:marecon}
For a radius parameter $\rho\in[0, \inr(\Omega)]$ and $g_{\rho}=\{x\in\MA(\Omega) \vert r(x)\geq \rho\}$,
\begin{equation}
    \Omega_{g_{\rho}}=\Omega \circ B_{\rho}.
\end{equation}{} 
\end{proposition}{}
For proof, see \S \ref{proof:medaxrecon}. The relationship between morphological operations and the morphological skeleton has been studied in image processing \cite{Lantuejoul1977}, where a similar statement is made in the discrete setting. Recall that the isoperimetric profile of no neck domains in $\mathbb{J}$ can be constructed by morphological opening. By Proposition \ref{prop:marecon} these can now be constructed from subsets $g$ of the medial axis. Furthermore, $g$ can be built iteratively due to the following result:
\begin{proposition}
\label{prop:hierarchic}
For generic $\Omega$ and $\rho_1 \geq \rho_2$, we have $g_{\rho_1} \subseteq g_{\rho_2}$. As extreme cases, $g_0=\MA(\Omega)$ and $g_{\inr(\Omega)} = \inx(\Omega)$.
\end{proposition}
\begin{proof}
If a point $x_1\in\R^2$ is in $g_{\rho_1}$, it must be on the medial axis and have $r(x_1)\geq\rho_1$. Then we also have $r(x_1)\geq\rho_2$, so it must be in $g_{\rho_2}$. At the smallest valid value $\rho=0$, $g_0$ is the entire medial axis, because the radius function is non-negative. At the largest nontrivial value $\rho=\inr(\Omega)$, $g_{\inr(\Omega)}$ only contains points at least $\inr(\Omega)$ away from the boundary. For generic $\Omega$, this is $\inx(\Omega)$ by definition.
\end{proof}{}

This motivates our algorithm in \S\ref{sec:algorithm} for constructing the isoperimetric profile by starting with $g_{\inr(\Omega)}$ and iteratively increasing $g$ by greedily absorbing neighboring medial axis points of highest radius. We will see in \S \ref{sec:alganalysis} why this medial axis based approach far outperforms direct computation of $\Omega \circ B_r$.

\vspace{-5px}
\subsection{Differential change in isoperimetric profile when traversing the medial axis}
Since our algorithm will traverse the medial axis, we consider here how much the isoperimetric profile of $\Omega_g$ will change as $g$ grows differentially. We can simplify this problem by looking at only a single medial axis edge segment $\gamma:[0,1]\rightarrow\R^2$, and let $g(\tau_2)=\bigcup_{\tau\in[\tau_1,\tau_2]} \gamma(\tau)$, for $\tau_1,\tau_2\in[0,1]$ and $\tau_1<\tau_2$. We derive that
\begin{equation}
\label{eq:differentialscore}
    \lim_{\tau_2\rightarrow \tau_1}\frac{\length(\partial \Omega_{g(\tau_2)}) - \length(\partial \Omega_{g(\tau_1)})}{\area(\Omega_{g(\tau_2)}) - \area(\Omega_{g(\tau_1)})} = \frac{1}{r(\gamma(\tau_1))}.
\end{equation}{}
For derivation see ``differentialGrowth.nb'' in supplemental materials. Equation \eqref{eq:differentialscore} tells us that if we infinitesimally increase $g$ by absorbing vertices of the medial axis near $x$, the slope of the isoperimetric profile will be $\nicefrac{1}{r(x)}$. A consequence of Equation \eqref{eq:differentialscore} is that the differential growth of the isoperimetric profile when growing a circle of fixed center from radius 0 is infinite, i.e., differentially, it is costly for $F$ in problem \eqref{eq:IP} to be disconnected. 

\vspace{-5px}
\section{Algorithm}
\label{sec:algorithm}

Our algorithm, as motivated by \S \ref{sec:medaxpartrecon}, can be roughly thought of as a greedy traversal of the medial axis. Given a domain $\Omega$, we  initialize $g = \{\inx(\Omega)\}$ and allow it to iteratively absorb adjacent points of the medial axis of maximal radius. 
As $|g|$ 
grows, $\area(\Omega_g)$ and $\length(\partial \Omega_g)$ form our IP upper bound.
To put this into practice, we must first discretize several quantities. 

First, we assume the input domain $\Omega$ is a polygon provided in the form of a $n\times 2$ vertex matrix $\overline \Omega$ (not a closure). 
Next we discretize the medial axis $\MA(\overline \Omega)$ into a graph $\MAG$ whose nodes densely sample the exact medial axis (details on this sampling below). Thus, $g$ is simply a subgraph of $\MAG$. 
Lastly, we discretize the limited reconstructions $F$ as polygons $\overline{F}=\overline{\Omega}_{g}$. 
This is computed as 
\begin{equation}
\label{eq:medrecon}
\overline{\Omega}_{g} = \bigcup_{e\in\text{$g$.edges}} \overline{\Omega}_{e},
\end{equation}
where each $\overline{\Omega}_{e}$ is the limited reconstruction from one edge of the medial axis graph. $\partial \overline{\Omega}_{e}$ consists of two semi-circular arcs and the linear subsets of $\partial \overline \Omega$ that are govenors of $e$. Therefore $\partial \overline{\Omega}_{g}$ is also
  a union of linear and circular arcs. We linearize circular arcs by polylines such that the length of any edge is less than input parameter $d_x$. As $d_x\rightarrow0$, $\overline{F}$ approaches $F$. 

The sampling of $\MA(\overline \Omega)$ into a graph $\MAG$ is performed with respect to two parameters $d_c$ and $r_l$. The \emph{radius limit} $r_l$ is the lower threshold radius for when we truncate the medial axis. In order to make sure we do not lose any information about necks of the domain, we always set $r_l \leq r_n$. In our experiments, $r_l=r_n$, unless we are on a no neck domain, in which case $r_l=$\delete{0}$.1$. For any choice of $r_l \leq r_n$, our algorithm can bound the profile for $t\in[0,\area(\Omega \circ B_{r_l})]$. Our discretization of the medial axis also discretizes our upper bound on the IP into a piecewise linear fit. 
If $e$ is a medial axis graph edge with endpoints $x_1$ and $x_2$, based on equation \eqref{eq:differentialscore}
we enforce that the difference between $\frac{1}{r(x_1)}$ and $\frac{1}{r(x_2)}$ is less than $d_c$. This concentrates samples of the IP upper bound to regions with high second derivative. As $d_c$ decreases, the resolution of our piecewise linear IP upper bound increases.

\begin{figure}
    \centering
    \includegraphics[width=.17\columnwidth]{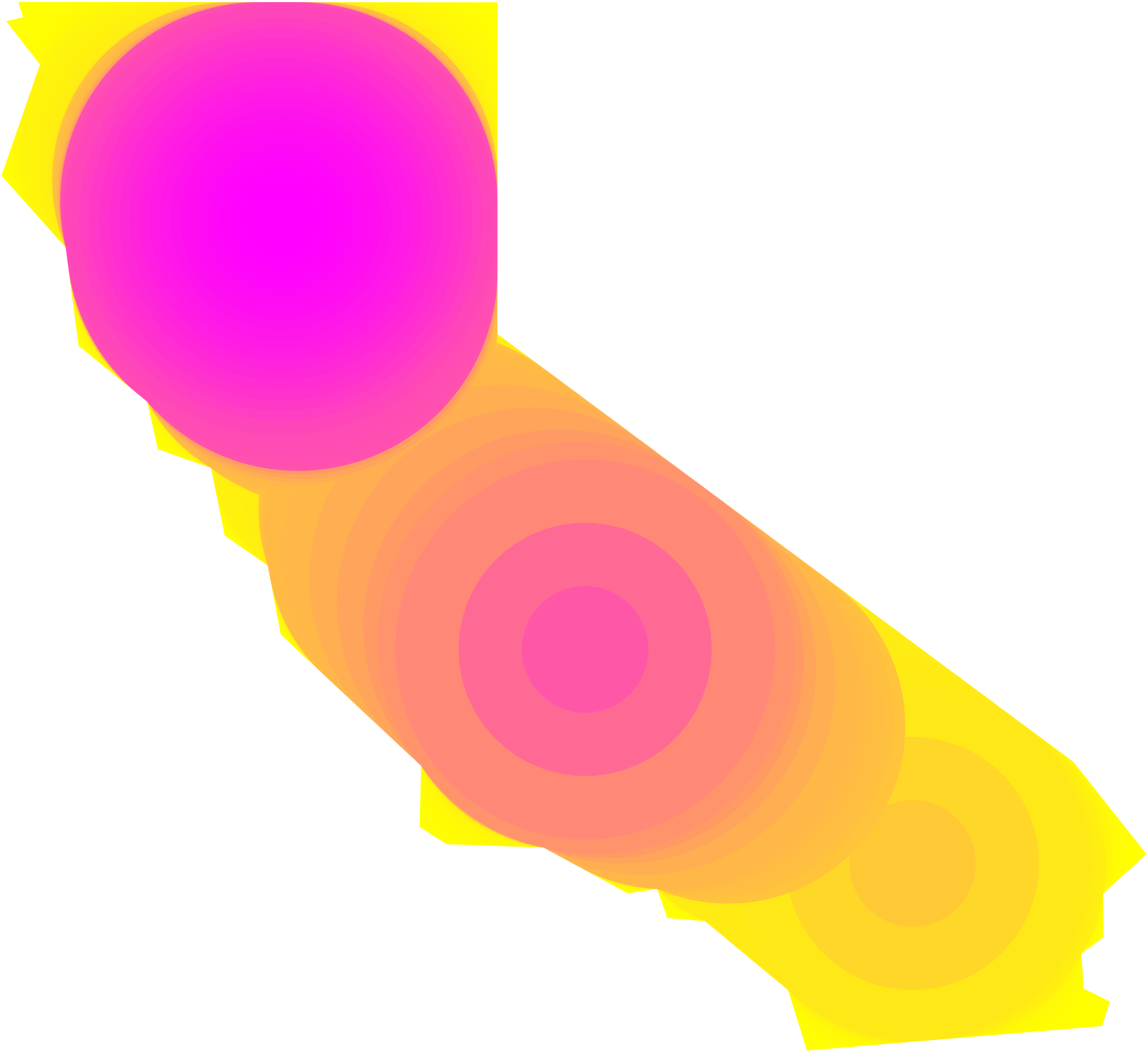}\hfill
    \includegraphics[width=.25\columnwidth]{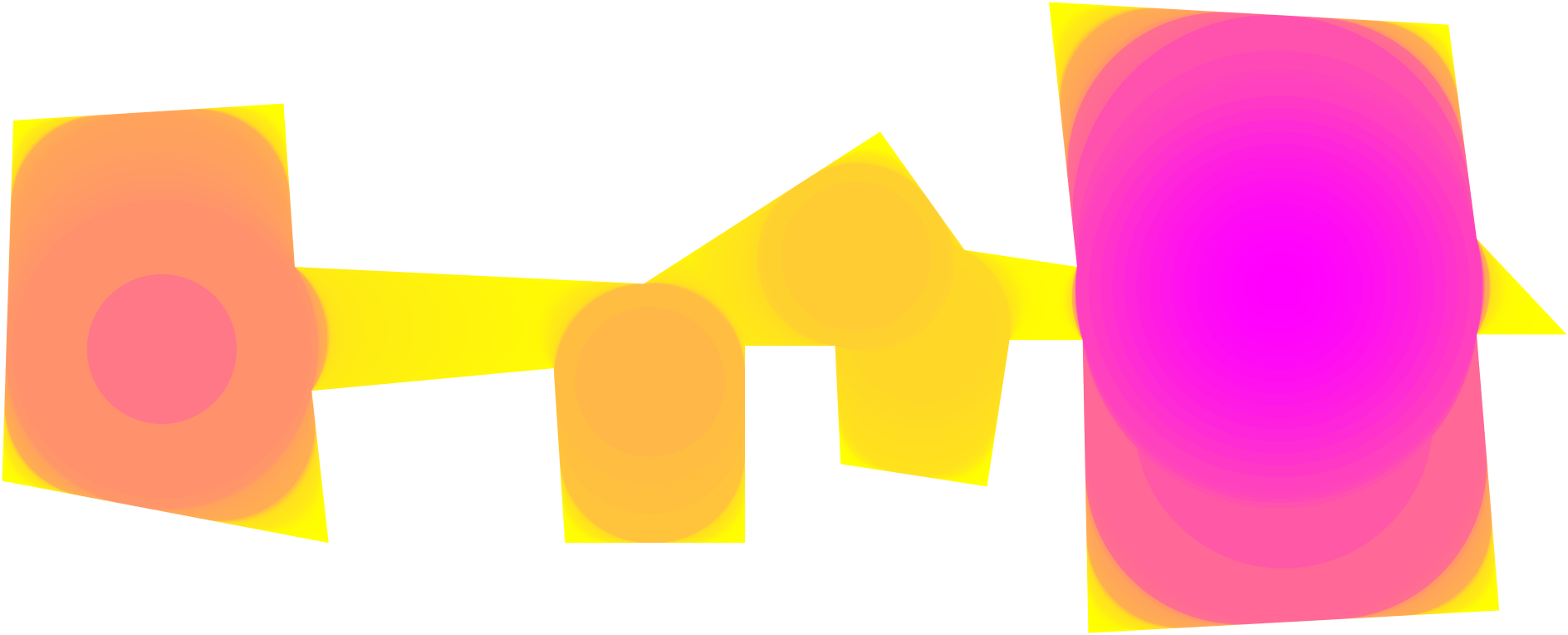}\hfill
    \includegraphics[width=.16\columnwidth]{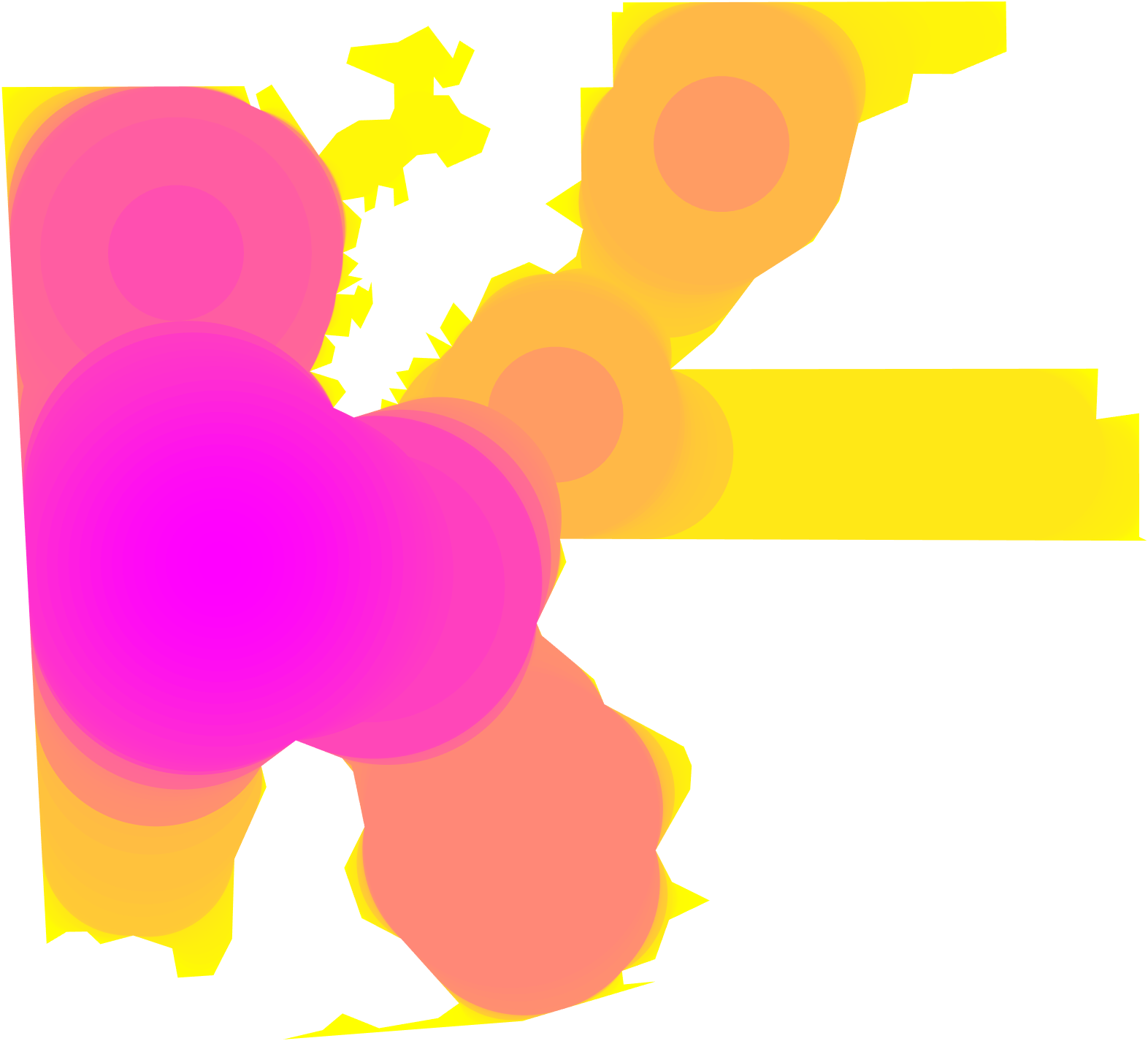}   \hfill \includegraphics[width=.14\columnwidth]{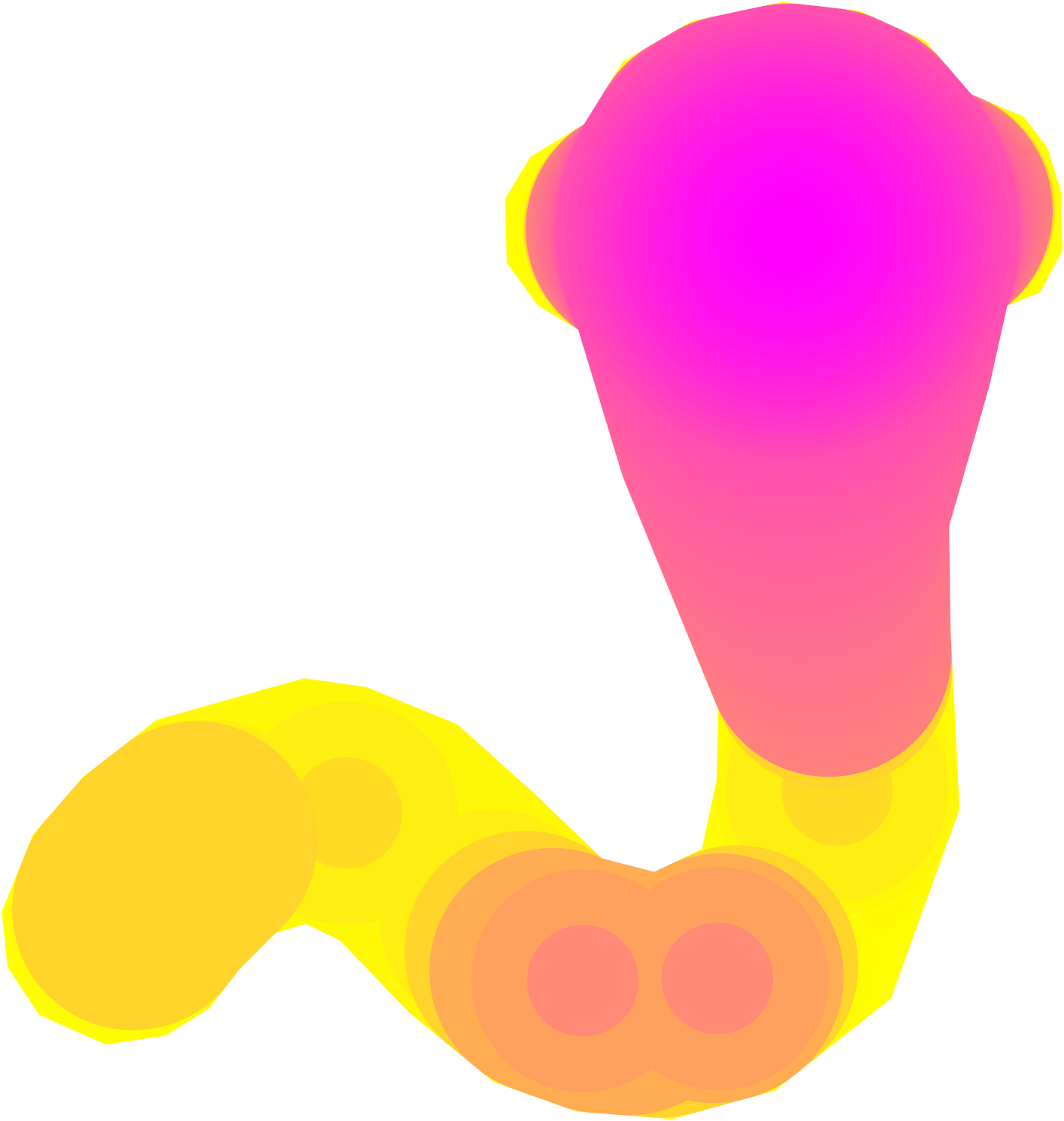}\hfill
    \includegraphics[width=.12\columnwidth]{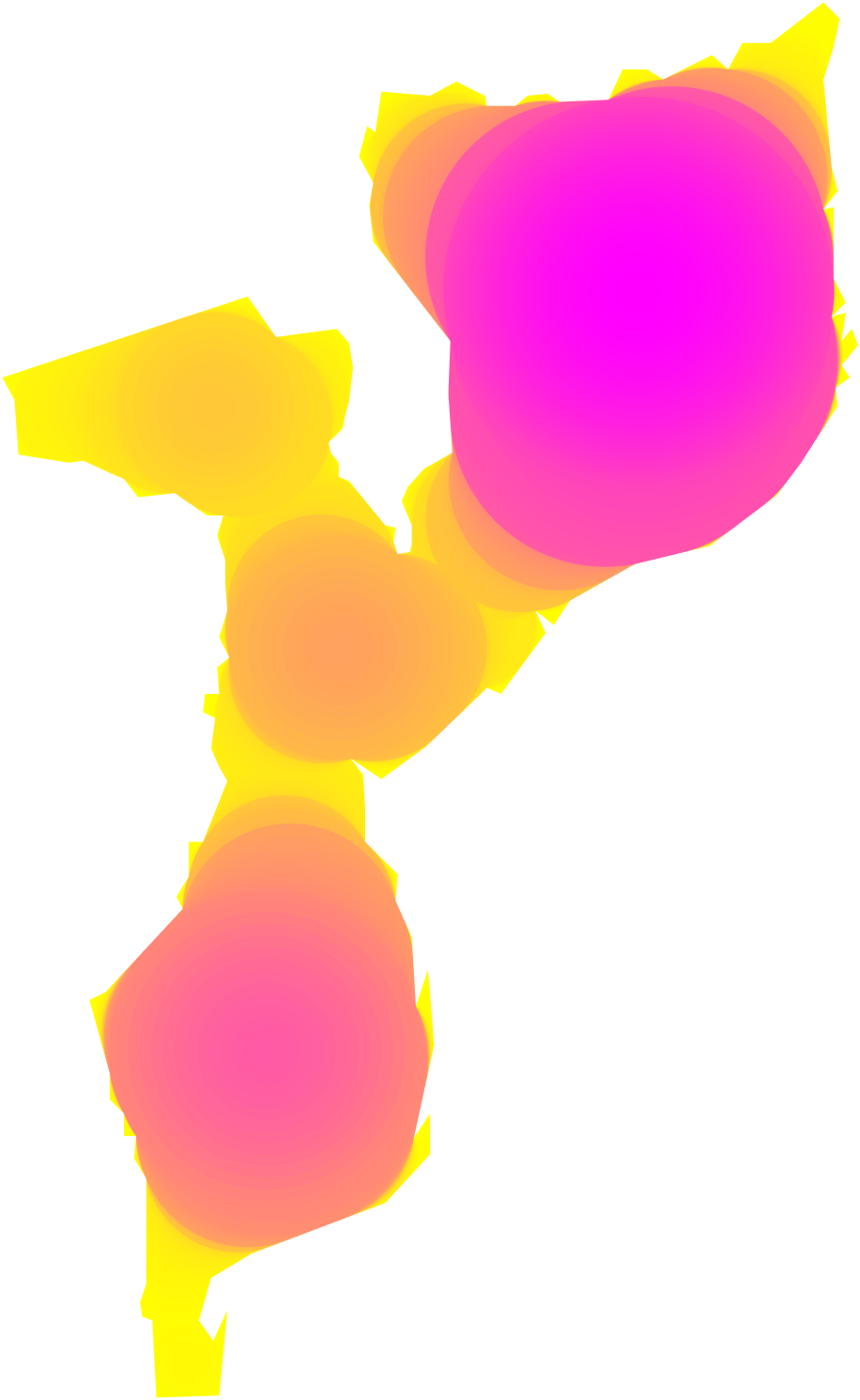} \\
    \includegraphics[width=.17\columnwidth]{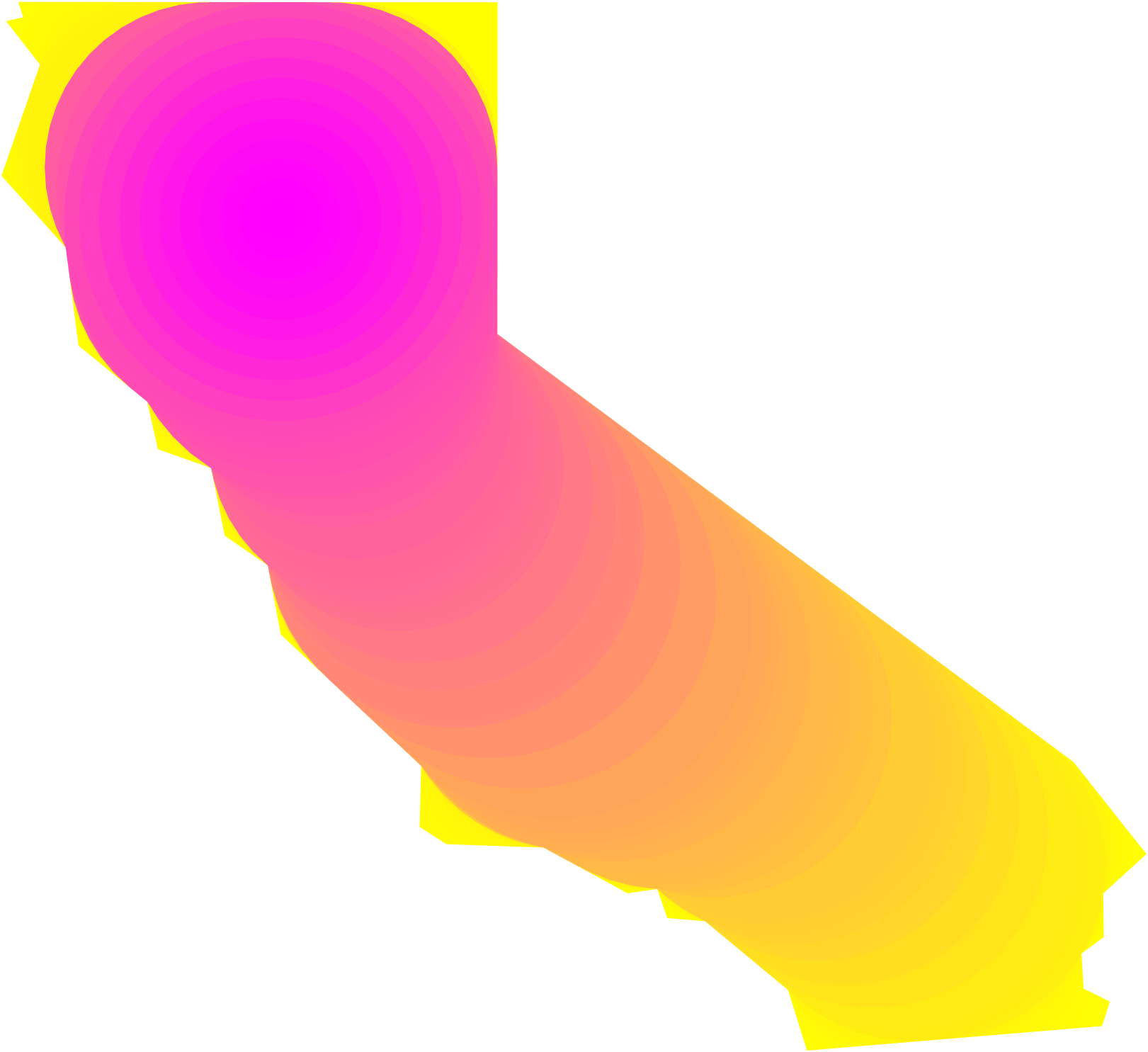}\hfill
    \includegraphics[width=.25\columnwidth]{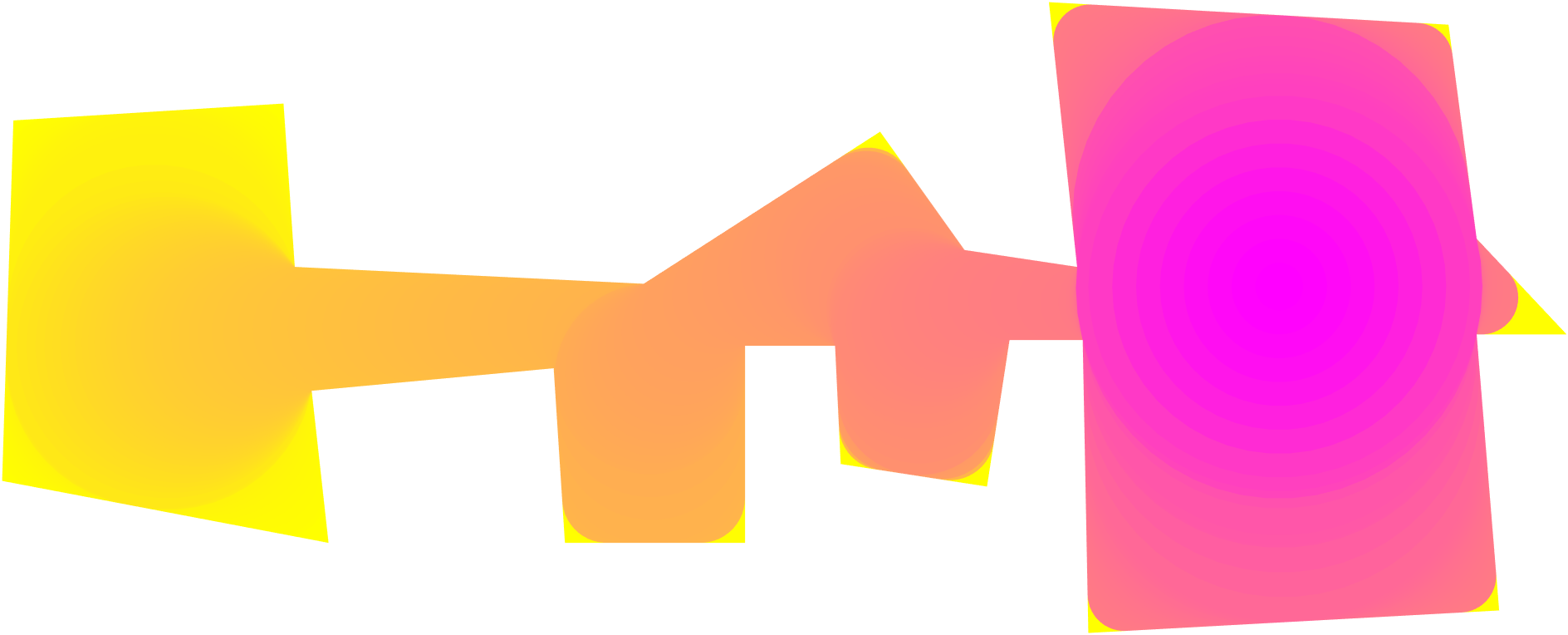}\hfill
    \includegraphics[width=.16\columnwidth]{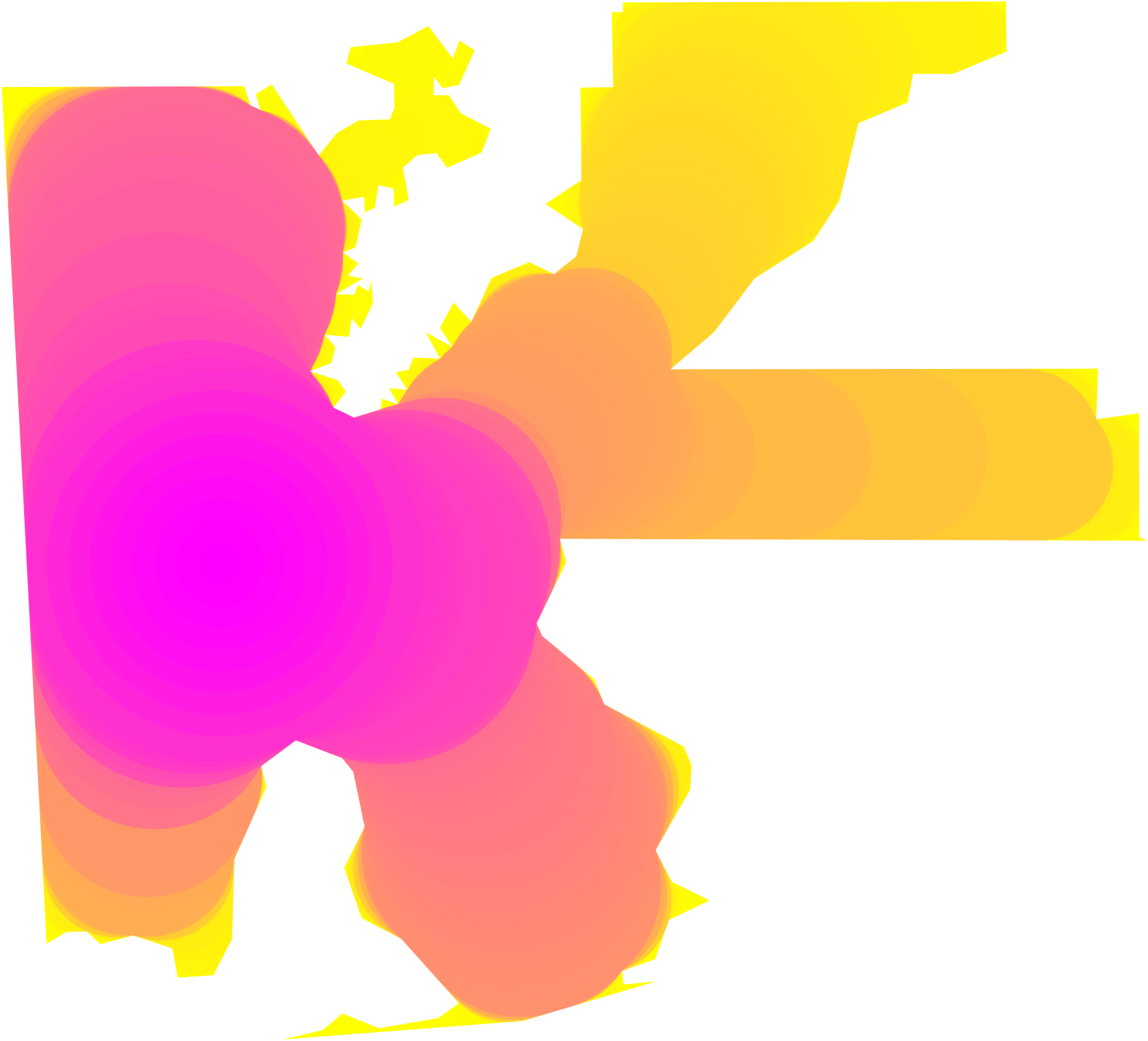}\hfill
    \includegraphics[width=.14\columnwidth]{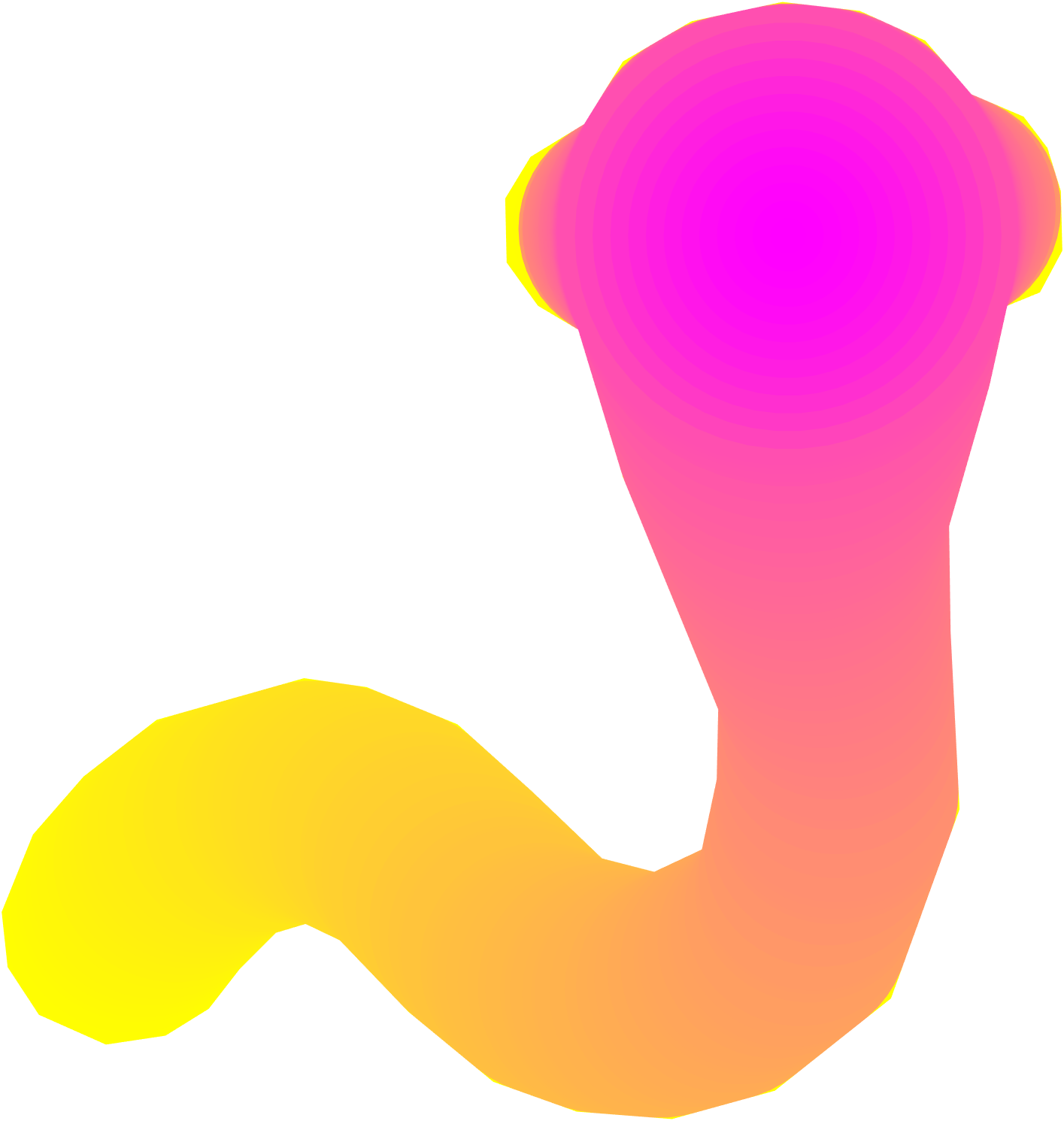}\hfill
    \includegraphics[width=.12\columnwidth]{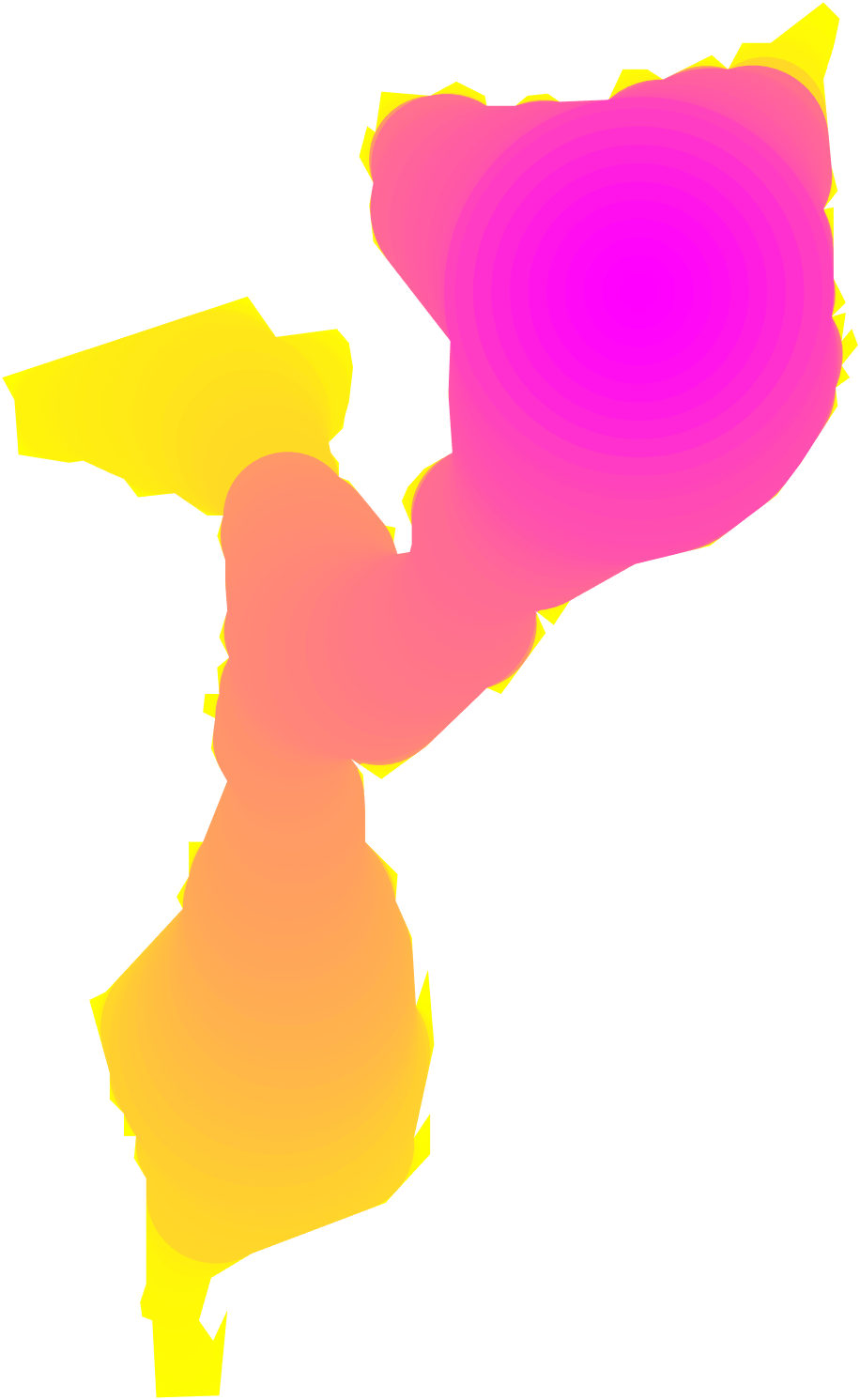}
    \caption{Sequence of inscribed subdomains $F\subset\Omega$ for \eqref{eq:IP} visualized as a colormap. From left to right the domains shown are \textbf{California}, \textbf{Boxes}, \textbf{District 1}, \textbf{Worm}, and \textbf{Mozambique}. Each point is colored by the earliest time on the IP bound when $F$ contained that point. The top row shows results from morphological opening \add{(with modification from \S \ref{sec:empanl})}, while the bottom row shows results from Algorithm \ref{alg:IPalg}.}
    \label{fig:Fsequence}
\end{figure}

Thus our algorithm is as follows. We first compute the exact medial axis of $\Omega$ using CGAL's edge Voronoi procedures \cite{cgalsegment}. Next, the medial axis is discretized into a graph $\MAG$ as described above. We then initialize $g^0=\{\inx(\Omega)\in\MAG\}$ and greedily absorb neighboring graph nodes and edges of maximal radius per iteration. In each iteration, $\overline{F_i}$ is the \delete{partial}\add{limited} reconstruction $\overline{\Omega}_{g^i}$. 
Our algorithm outputs the sequence of $\overline{F_i}$, their areas $t_i$ and their perimeters $p_i$.
This procedure is summarized by Algorithm \ref{alg:IPalg}. Figure \ref{fig:Fsequence} provides a visualization of the $\overline{F_i}$ constructed by this procedure.

\vspace{-10px}
\subsection{Direct $\Omega \circ B_r$}
For comparison with the direct computation of $\Omega \circ B_r$ suggested in \S \ref{sec:morphopen}, we pick 100 evenly spaced values of $r_i\in[0,\inr(\Omega)]$, compute $\Omega \circ B_{r_i}$ and its corresponding area and perimeter. This is easily implemented with Matlab's \texttt{polybuffer} methods and forms a second upper bound that in the majority of test cases performs worse than our medial axis based approach. In very few cases, it can perform better, in which case our final IP upper bound is the minimum of these two upper bounds.


{}

\begin{algorithm}[t]
\begin{algorithmic}[1]
 \Procedure{Compute-IP-Bound}{$\overline{\Omega}$, $d_c$, $r_l$}
 \State $[\MA(\overline{\Omega}), r(\cdot)]\gets$ \Call{GetMedialAxis}{$\overline{\Omega}$}
 \State $\MAG\gets$ \Call{MAtoGraph}{$\MA(\overline{\Omega})$, $d_c$, $r_l$}
 \State $g^0 \gets \inx(\overline{\Omega})$
 \State $\overline{F}_0 \gets \inB(\overline{\Omega})$
 \State $t_0\gets \area(\overline{F_0})$
 \State $p_0\gets \length(\partial \overline{F_0})$
 \While{$g^i\neq\MAG$}
    \State $v\gets$\Call{GetAdjacentNodes}{$\MAG$, $g^i$.nodes}
    \State $c\gets v \backslash g^i$.nodes
    \State $c^*\gets$\Call{ArgMax}{$r(c)$}
    \State $e^*\gets$\Call{GetEdgeBetween}{$c^*$, $g^i$}
    \State $g^{i+1} \gets$ \Call{AddNode}{$g^{i}$, $c^*$}
    \State $g^{i+1} \gets$ \Call{AddEdge}{$g^{i+1}$, $e^*$}
    \State $\overline{F_{i+1}} \gets \overline{\Omega}_{g^{i+1}}$
    \State $t_{i+1} \gets \area(\overline{F_{i+1}})$
    \State $p_{i+1} \gets \length(\partial \overline{F_{i+1}})$
 \EndWhile
 \State \Return $t,p,\overline{F}$
 \EndProcedure
\end{algorithmic}
 \caption{Isoperimetric Profile Upper Bounder}
 \label{alg:IPalg}
\end{algorithm}

\vspace{-10px}
\section{Theoretical Tools}
Our algorithm in the previous section produces a sequence of inscribed shapes $\overline{F}$. We woud like to analyze how close they are to optimal. To do that, we first establish some theoretical tools that will help us analyze the behavior of our algorithm. These tools bridge the gap between the definition of necks and the medial axis.

\vspace{-10px}
\subsection{Medial Axis Necks}
\label{sec:medaxisnecks}
\begin{figure}
    \centering
    \includegraphics[width=0.49\columnwidth]{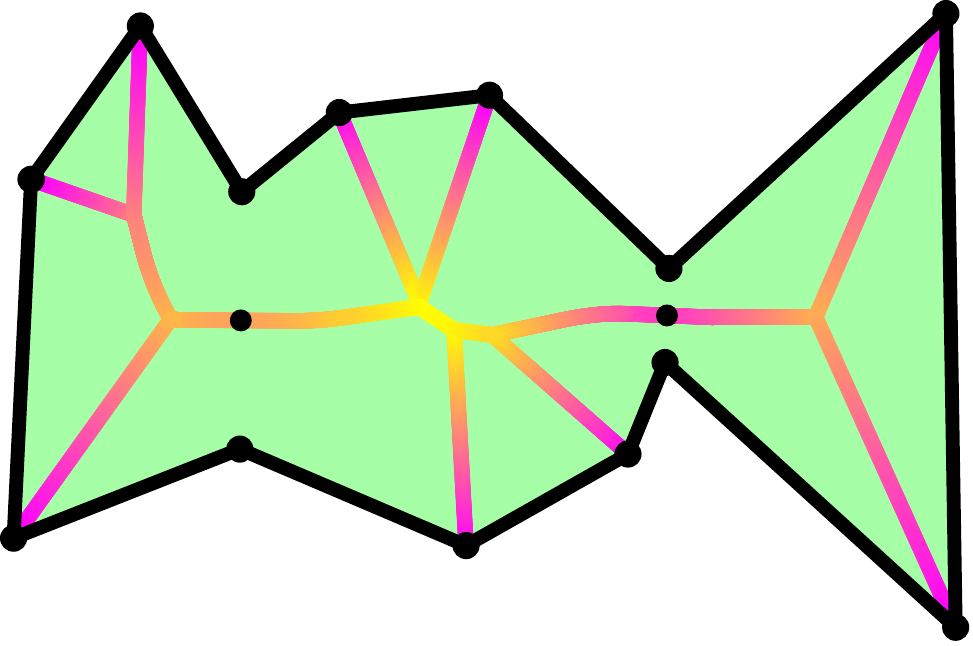}\hfill
    \includegraphics[width=0.415\columnwidth]{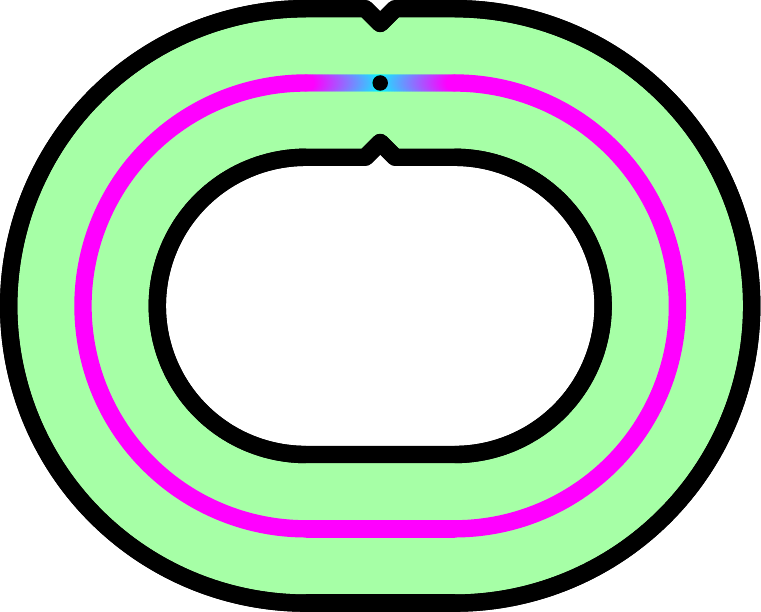}
    \caption{Domains with medial axes indicated by colored curves and and necks depicted by isolated black points. (Left) Domain containing 2 necks of different sizes (Right) Pinched annulus with one neck by definition \ref{def:neck2}, but no necks by definition \ref{def:neck}}
    \label{fig:neckdemo}
\end{figure}

Since our algorithm is derived jointly from the medial axis and the concept of necks, we create a generalized definition for necks based on the medial axis transform. From this definition, we can quantify the size of multiple necks as well as their locations (see Figure \ref{fig:neckdemo}). This will allow us to prove results that are not possible to state with Definition \ref{def:neck}.
\begin{definition}[Generalized Necks]
\label{def:neck2}
$\eta(\Omega)$ is the set of $x\in\MA(\Omega)$ that satisfy either of the following:
\begin{enumerate}
    \item $x$ is on the interior of a medial axis edge segment, and $r(x)$ is a local minimum at $x$.
    \item $x$ is on a medial axis node and there are two distinct outgoing medial axis edge segments $e_1$, $e_2$ on which the change in $r(x)$ in both of those directions is positive.
\end{enumerate}
The size of a neck $x\in\eta(\Omega)$ is $r(x)$.
\end{definition}

{}

Algorithmically, the computation of the necks $\eta(\Omega)$ is done by first computing the medial axis, and then iterating over its edges and nodes while checking the neck conditions.
We denote the maximum neck radius by
$r_m = \max\{r(x): x\in\eta(\Omega)\}$. Similarly we denote its minimum neck radius by
$r_n = \min\{r(x): x\in\eta(\Omega)\}$. If a domain has no necks then we say $r_n=\infty=r_m$. The maximum neck radius $r_m$ is not part of the previous definition of necks but will be used in Proposition \ref{prop:generaldomain} to give theoretical guarantees about our algorithm. A nice property of our neck definition is that it is compatible with the previous neck definition, i.e., the minimal neck radius of both are identical.

\begin{proposition}[Equality of necks]
\label{prop:neckequality}
For $\Omega \in \mathbb{J}$, $r_n$ of Definition $\ref{def:neck}$ is equivalent to $r_n$ of Definition $\ref{def:neck2}$.
\end{proposition}{}
For proof see \S \ref{proof:neckneck}. Note that this equivalence does not hold for domains outside $\mathbb{J}$. Consider the annulus with a pinch in it, illustrated in Figure \ref{fig:neckdemo}. This does not have a neck by Definition \ref{def:neck}, but by Definition \ref{def:neck2} it has a single neck marked by the black point. 

\vspace{-10px}
\subsection{Thick Necks}
While much is known about the isoperimetric profile for shapes with no necks ($r_n=\infty$), much less is known when $r_n$ is finite. One might hope that if $r_n$ were in some sense \emph{large enough}, we could still compute the isoperimetric profile with some confidence. 

Here we define the \emph{thick neck} condition which will be used in Proposition \ref{prop:thickneck} to analyze the behavior of our algorithm on domains whose necks are large enough.
Recall that $\inB(\Omega)$ is the largest circle inscribed in $\Omega$.
\begin{definition}[Second Largest Circle]
Let the radius of the second largest circle in $\Omega$ without replacement be 
\begin{equation}
    \inr_2(\Omega) = \inr(\Omega - \inB(\Omega)).
\end{equation}{}
\end{definition}{}
\begin{definition}[Thick Neck]
A neck $x\in\eta(\Omega)$ is \emph{thick} if 
\begin{equation}
    r(x) > \frac{\inr_2(\Omega)}{2}
\end{equation}{}
\end{definition}{}

\begin{definition}[Thick Neck Condition]
A domain satisfies the \emph{thick neck} condition if all of its necks are thick.
\end{definition}{}
For examples of domains that satisfy the thick neck condition and their profiles, see Figure \ref{fig:thicknecks}. 

\vspace{-5px}
\section{Analysis of Algorithm \ref{alg:IPalg}}
\label{sec:alganalysis}
Given the tools of the previous section, we can now analyze how the perimeter of our constructed shapes $\overline{F_i}$ compares to that of $\partial E_\Omega(t)$. At the $t_i$’s where our algorithm samples $\overline{F_i}$, and $p_i$, our bound satisfies several properties.

\vspace{-10px}
\subsection{Theoretical analysis}

By construction, Algorithm \ref{alg:IPalg} produces an upper bound to the isoperimetric profile. This is because $\overline{F_i}$ is always a feasible candidate to \eqref{eq:IP} with $t_i =\area(\overline{F_i})$. By applying Equation \eqref{eq:differentialscore} to domains in $\mathbb{J}$, our bound is strictly monotonically increasing, similarly to the actual IP. Our construction also satisfies the condition in \cite{gonzalez1980regularity}, which states that $\partial E_{\Omega}(t)$ must lie tangent to $\partial \Omega$ and \add{its free boundary must} have no points of singular curvature. \add{This is because the free boundary of $\overline{F_i}$ is made of circular arcs that start and end on $\partial \Omega$ and whose centers are on different edges of the medial axis. If these circular arcs intersect to form a point of singular curvature, then a non-contractable loop can be drawn in $\Omega$ by tracing from the intersection through both edges of the medial axis, back to $\inx(\Omega)$, implying $\Omega \notin \mathbb{J}$.}

We provide guarantees for when our bound is tight for any domain in $\mathbb{J}$. While we cannot guarantee that our bound is tight for all $t_i$, by Proposition \ref{prop:generaldomain}, we know our bound is tight for $t_i$ in a limited range.
\begin{proposition}
\label{prop:generaldomain}
Let 
$\Gamma_r$ be the connected component of $\Omega \ominus B_r$ that contains $\inx(\Omega)$.
Let $r_t=\max(r_m, \nicefrac{\inr_2(\Omega)}{2})$.
For any domain in $\mathbb{J}$, our bound for $IP_{\Omega}(t_i)$ is tight for $t_i > \area(\Omega \circ B_{r_n})$ and $t_i < \area(\Gamma_{r_t} \oplus B_{r_t})$.
\end{proposition}{}
\begin{proof}
For $t_i > \area(\Omega \circ B_{r_n})$, we know that $E_{\Omega}(t_i)=\Omega \circ B_{r_n}$. In this region, the remaining area in $\Omega$ not covered by $E_{\Omega}(t_i)$ has no more necks. Thus the problem reduces to the case with no necks, on which our algorithm produces tight bounds. 
For proof of the latter range of $t_i$ values, see Appendix \ref{proof:earlyT}.
\end{proof}

While Proposition \ref{prop:generaldomain} does not cover the entire IP, it applies to any domain in $\mathbb{J}$ regardless of the presence or size of necks. To achieve this result, this proposition uses $r_m$, a quantity that is not defined in prior work. 

\begin{figure}
    \centering
    \includegraphics[width=1\columnwidth]{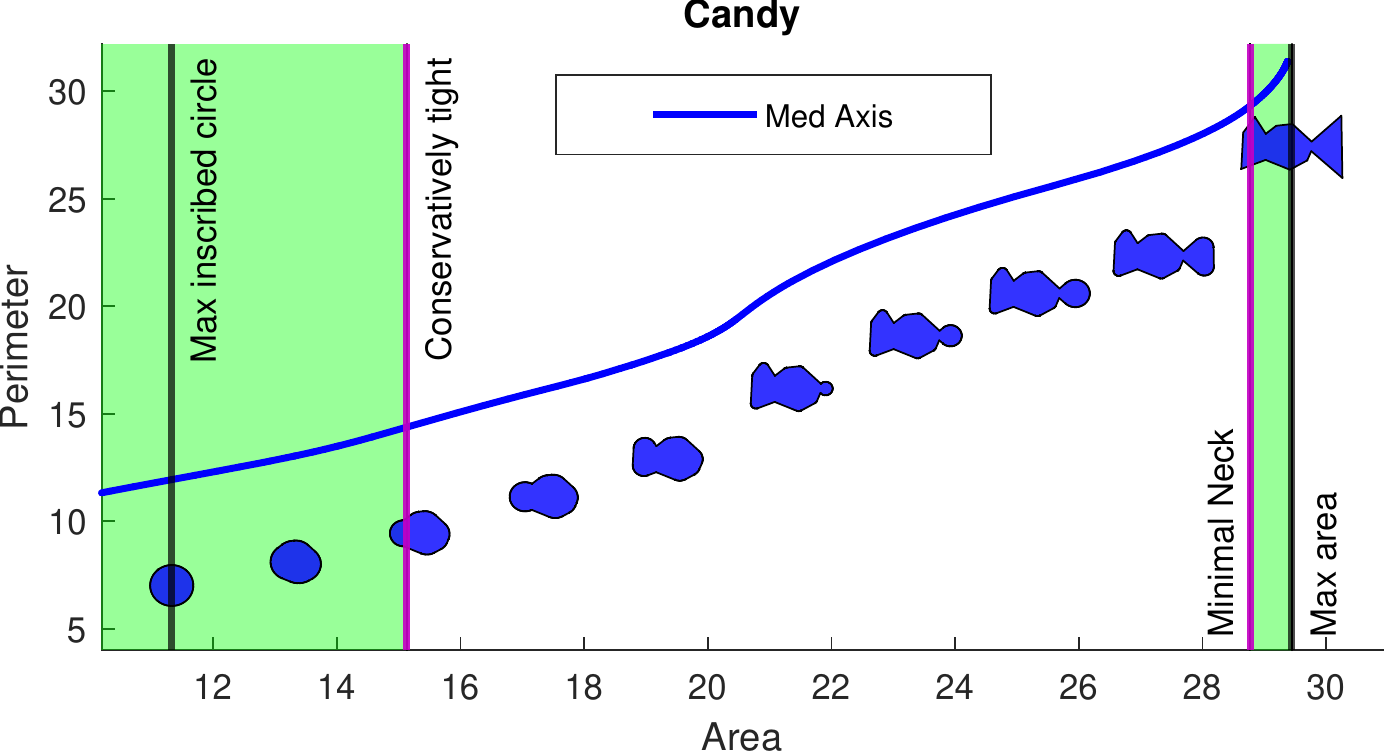}
    \caption{Bounds on the IP computed by our medial axis traversal algorithm \add{on the \textbf{Candy} domain}. By Proposition \ref{prop:generaldomain} our bound is tight on the green regions (left of ``Conservatively tight'' and right of ``Minimal Neck''). On the rest of the profile, our medial axis traversal algorithm is an upper bound. Visualizations of $F_i = \Omega_g$ are depicted along the profile bound. }
    \label{fig:morphopenvma}
\end{figure}

Application of Proposition \ref{prop:generaldomain} is demonstrated in Figure \ref{fig:morphopenvma}. We compute our upper bound to the IP by Algorithm \ref{alg:IPalg} on the \textbf{Candy} domain. Its profile is depicted between two events: the area of the ``max inscribed circle'' $\area(\inB(\Omega))$, and the ``max area'' $\area(\Omega)$. We apply Proposition \ref{prop:generaldomain}, which tells us that our medial axis bound for the IP is tight to the left of the ``Conservatively tight'' purple event line and to the right of the ``Minimal Neck'' purple event line. For values of $t_i$ between the two purple lines, we upper-bound the IP.

{}

Given some assumptions on the domain $\Omega$, we show that our IP bound exactly recovers the IP.
\begin{proposition}
\label{prop:thickneck}
If $E_{\Omega}(t)$ grows only continuously except when transitioning to being disconnected, 
then for $\Omega\in\mathbb{J}$ that satisfy the thick neck condition, our bound is tight for all $t_i$.
\end{proposition}
\add{Intuitively, this is because the smallest perimeter cost for having a disconnected $E_{\Omega}(t)$ is larger than the cost for continuing to grow $E_{\Omega}(t)$ via limited reconstructions.}
For proof, see \S \ref{proof:bigneck}. Proposition \ref{prop:thickneck} allows us to compute the exact isoperimetric profile for domains satisfying the thick neck condition. We show in \S \ref{sec:results:thickneck} examples of domains satisfying this condition and that while previous methods of computing the isoperimetric profile are not tight, our method provides the exact profile for this larger class of domains.

\vspace{-10px}
\subsection{Empirical Quality Improvements}
\label{sec:empanl}

\begin{figure}
    \centering
    \includegraphics[width=.9\columnwidth]{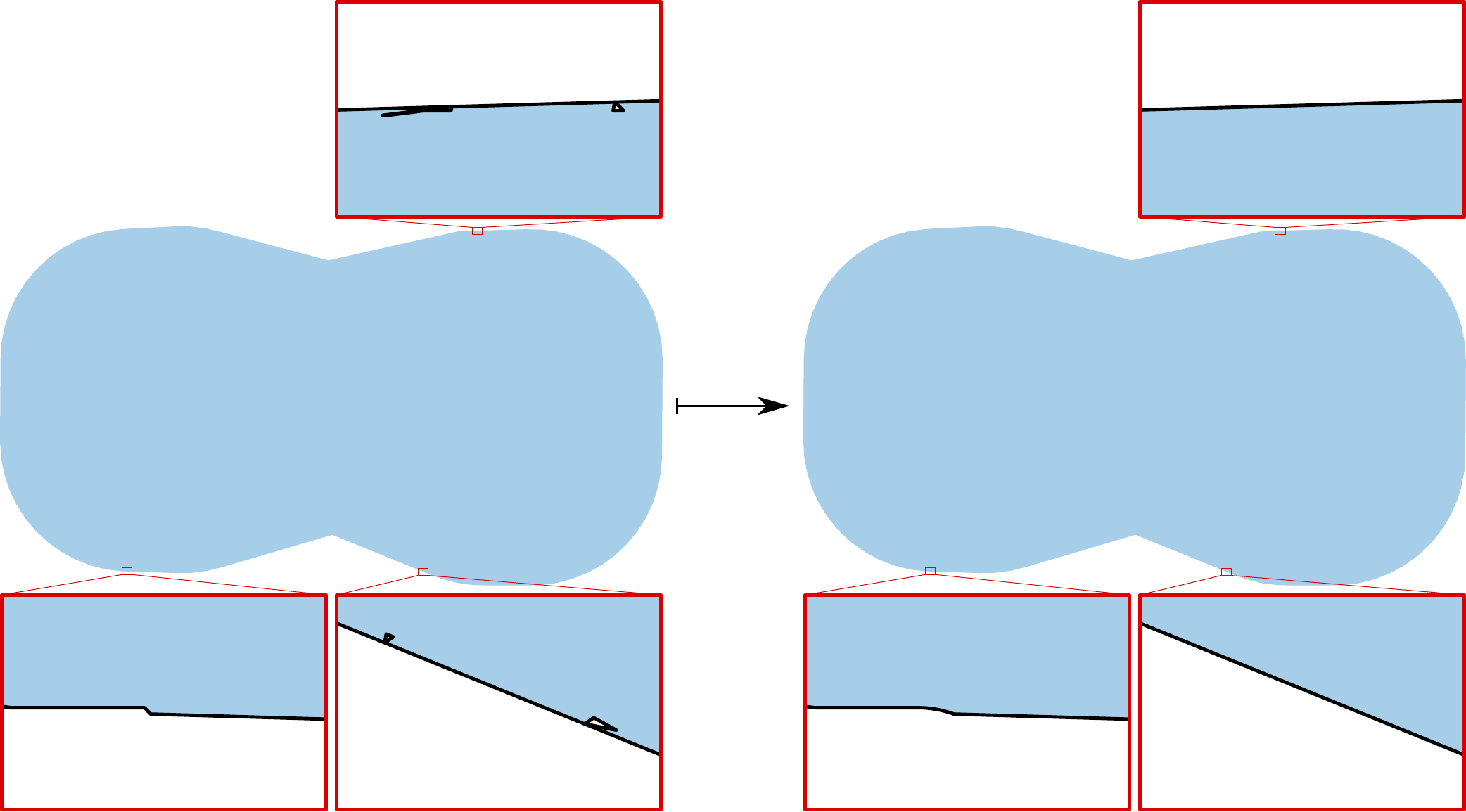}
    \caption{(Left) $\overline{F}_i$ before post processing. Several holes and boundary flaws are visible by magnification. (Right)  $\overline{F}_i$ with post processing. Holes and boundary flaws are repaired. Post processing changes the area by less than 1e-5, while decreasing the perimeter by 0.1.}
    \label{fig:mopenfix}
\end{figure}

\begin{figure}
    \centering
    \includegraphics[width=.49\columnwidth]{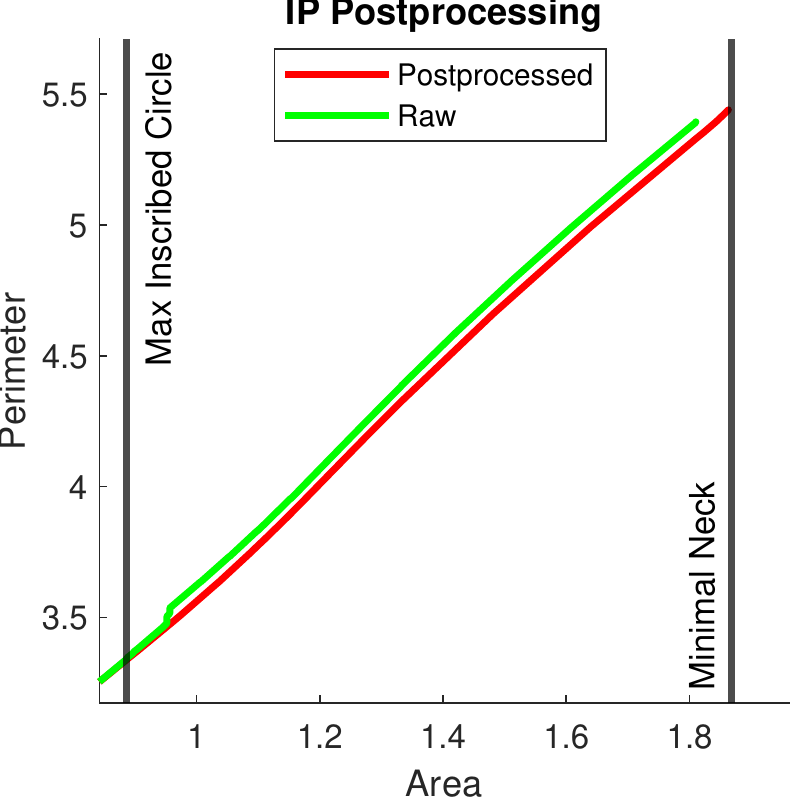}
    \includegraphics[width=.49\columnwidth]{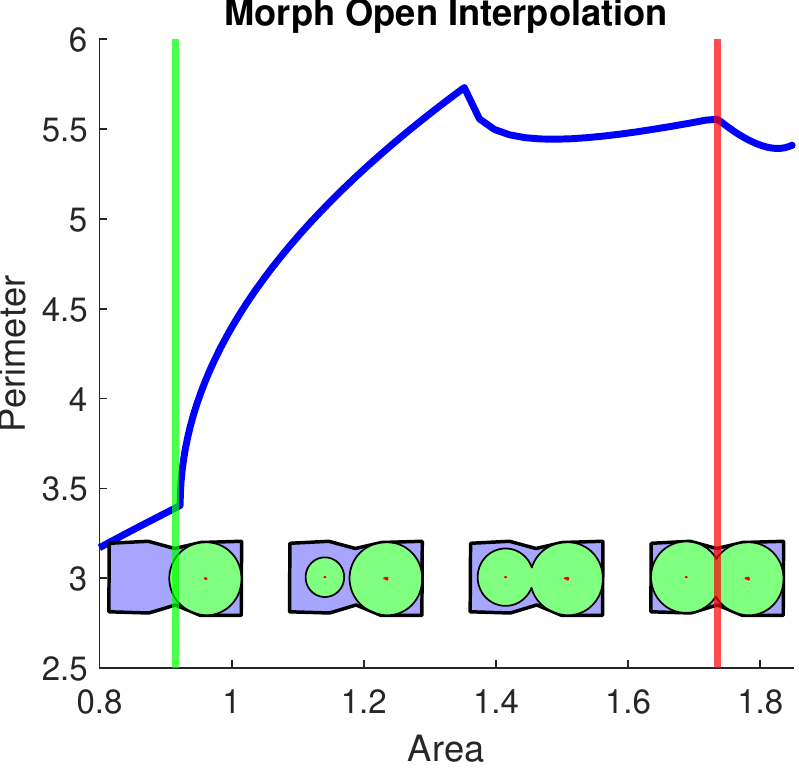}
    \caption{(Left) Our post processing procedure gets rid of small floating point artifacts in the inscribed shape resulting in lower perimeter and a tighter IP bound. (Right) The profile generated by $\Omega\circ B_r$ disconnectedly jumps from the green event line to the red event line. $\Omega\ominus B_r$ is visualized in red and shows the creation of a very small disconnected region at $x^-$. We interpolate the profile in between these discontinuous transitions by inflating a circle at $x^-$ from radius $0$ to $r$.}
    \label{fig:postprocess}
\end{figure}{}

Following Equation \eqref{eq:medrecon} we compute $\Omega_g$ by the union of floating precision polygons. This sometimes introduces artifacts like holes, or jagged boundary at a scale that is invisible. While seemingly negligible, many holes or jagged regions add up to significantly increase the perimeter thus loosening our upper bound. This is solved easily by some basic post processing. First we remove all holes in $F$, then we perform morphological closing with a ball of small radius, and lastly we intersect it with $\Omega$ to make it feasible again. These steps are summarized in Algorithm \ref{alg:polypost}. Figure \ref{fig:mopenfix} shows the change in $\overline{F}_i$ before and after post processing. Figure \ref{fig:postprocess} demonstrates the difference in the IP with and without post processing. 

We also perform a repair step to the upper bound obtained by direct computation of $\Omega \circ B_r$. Since $\Omega \circ B_r$ changes discontinuously with $r$ for domains with necks, this leaves gaps in the IP. An example of this discontinuous transition is visualized in Figure \ref{fig:postprocess}. The discontinuous change happens due to the creation of a new disconnected point $x^-$ in $\Omega \ominus B_r$. $\Omega \circ B_r$ will instantly include a circle centered on $x^-$ of radius $r$, which abruptly increases area and perimeter. We detect when the number of disconnected regions in $\Omega \ominus B_r$ changes, and compute $x^-$ as the center of the bounding box of the smallest area component of $\Omega \ominus B_r$. We then inflate a disk centered at $x^-$ from radius $0$ to $r$. This procedure continuously fills in the IP bound and is illustrated in Figure \ref{fig:postprocess}. 

\begin{algorithm}[t]
\begin{algorithmic}[1]
 \Procedure{PostProcessPolygons}{$F$, $r_c$, $\Omega$}
 \State $F \gets$\Call{RemoveHoles}{F}
 \State $F \gets F\bullet B_{r_c}$
 \State $F \gets F \bigcap \Omega$
 \State \Return $F$
 \EndProcedure
\end{algorithmic}
 \caption{PostProcess Polygon}
 \label{alg:polypost}
\end{algorithm}

\begin{figure}
    \centering
    \includegraphics[width=1\columnwidth]{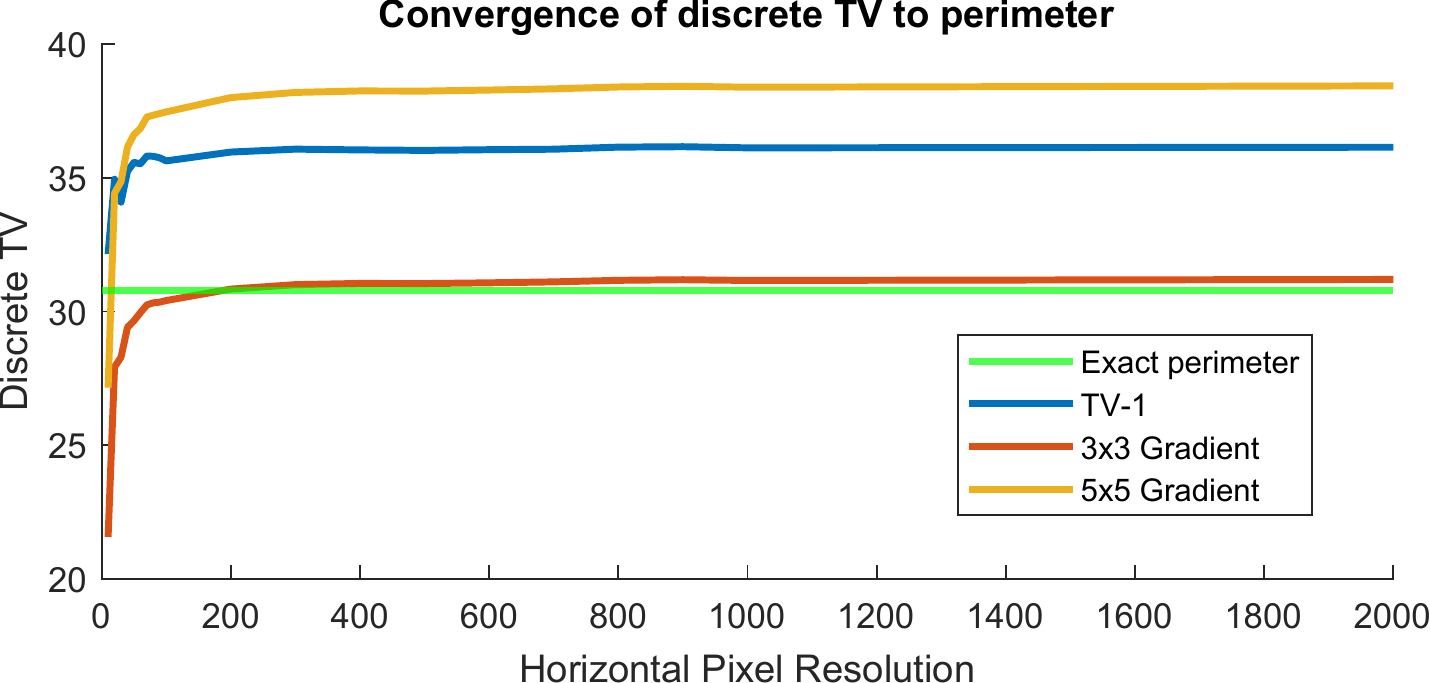}\\
    \includegraphics[width=.15\columnwidth]{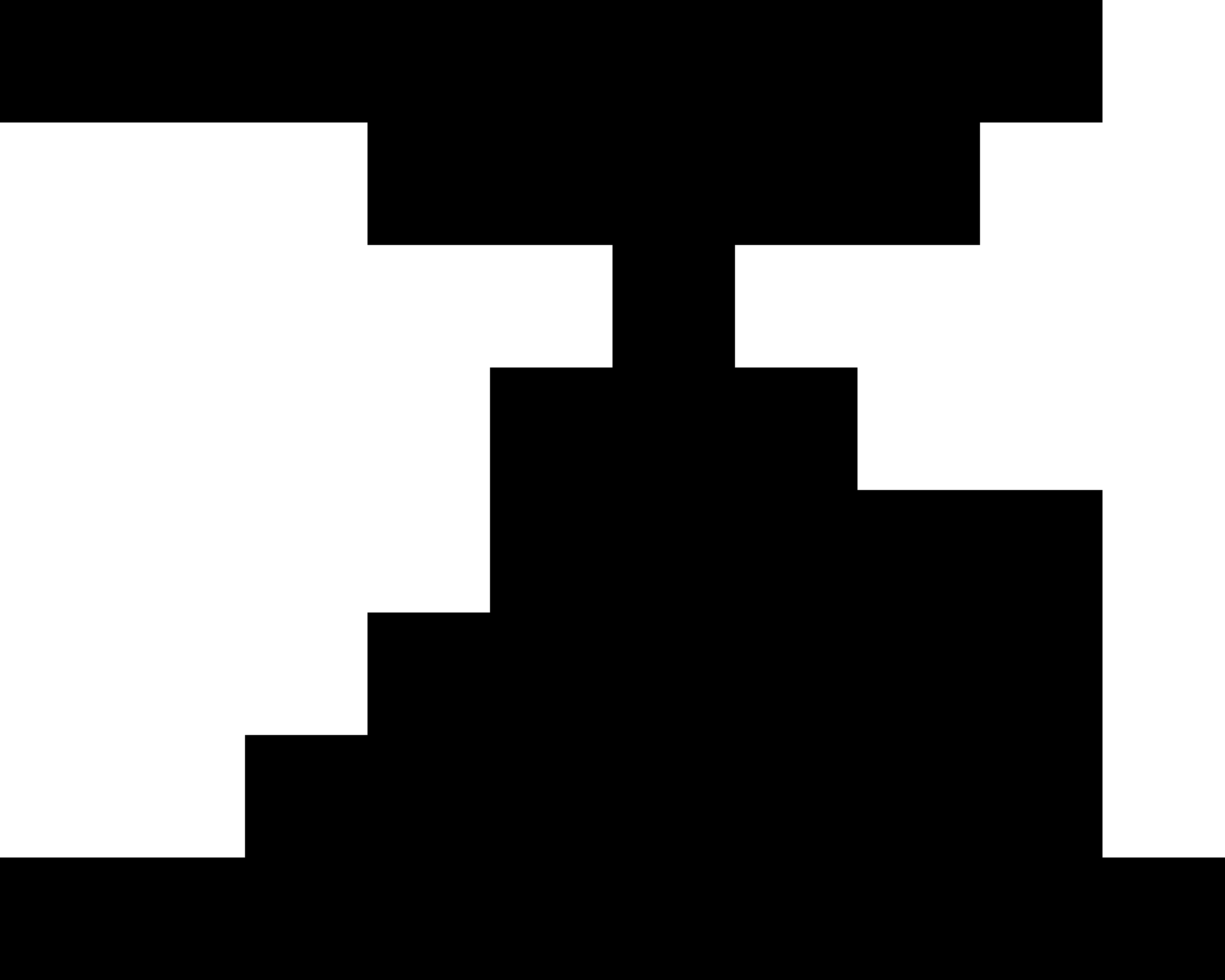}
    \hfill
    \includegraphics[width=.1875\columnwidth]{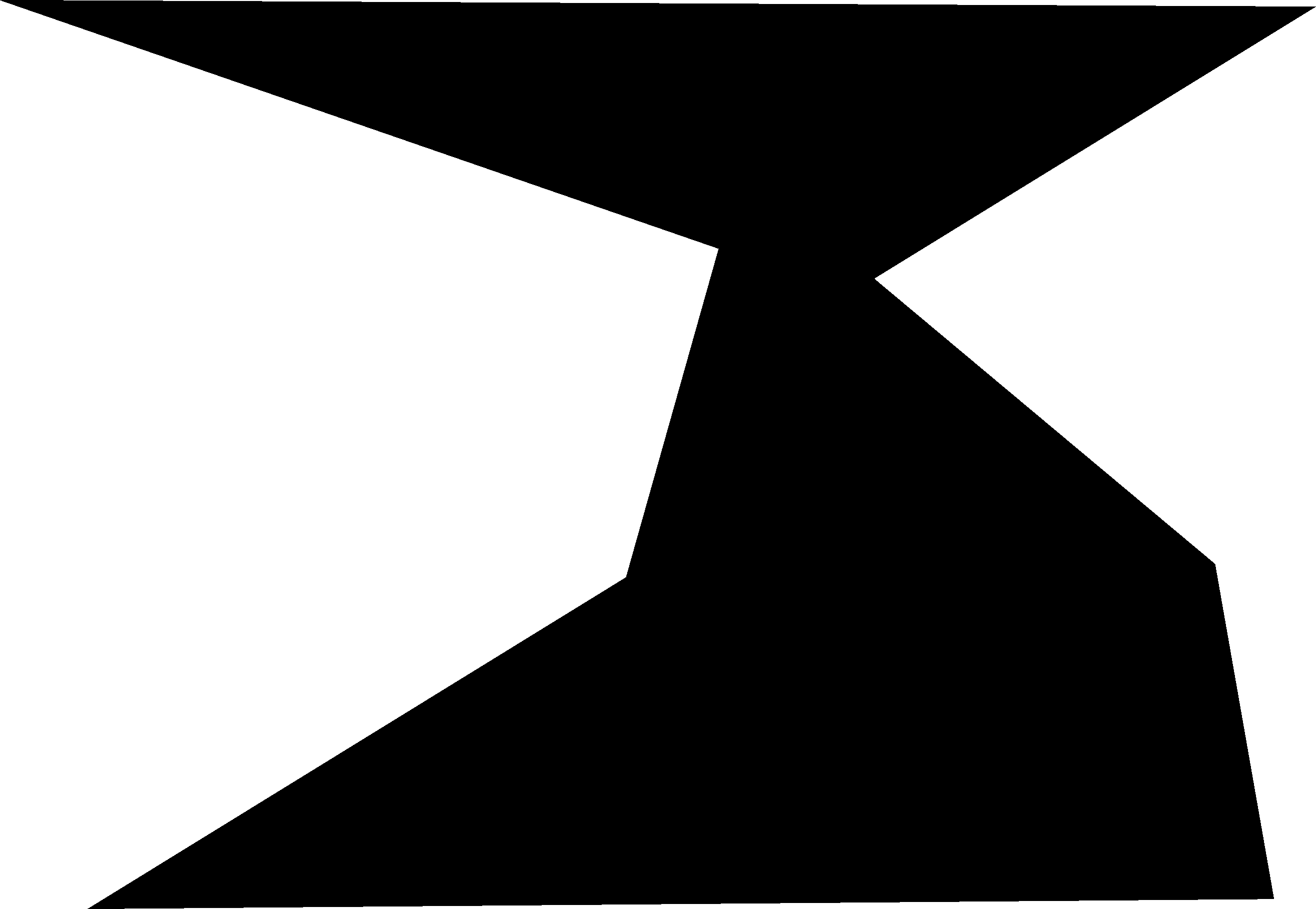}
    \caption{We test three different discretizations of total variation (TV). (1) TV-1: the TV discretization of \cite{DeFord2019TotalVI}. (2) 3x3 Gradient: TV where gradients are computed with a 3x3 stencil. (3) 5x5 Gradient: TV where gradients are computed with a 5x5 stencil. We test how well these three discretization perform at approximating the perimeter of a polygon when applied to an indicator function of it on finer grid resolutions. While none of the three converge exactly to the correct perimeter, the 3x3 gradient performs significantly better.}
    \label{fig:tvdisc}
\end{figure}{}

\vspace{-10px}
\section{Experiments and Results}
\label{sec:results}

\subsection{Comparison Methods}
We compare our algorithm against the few existing computational approaches to the isoperimetric profile. First we compare Algorithm \ref{alg:IPalg} with direct morphological opening. While both generate upper bounds, direct opening has much fewer guarantees and generally produces a looser bound.
Next we compare with the TV approach \cite{DeFord2019TotalVI} which generates lower bounds. In theory, it generates the convex lower envelope of the IP. 

Code for our algorithm, figure generation, and all parameters used can be found in supplementary materials. We show profiles starting from the area of the maximum inscribed circle because the profile before then is uninformative.

\vspace{-5px}
\subsubsection{Modification to comparison method}
For comparisons with TV \cite{DeFord2019TotalVI} we use a modification of their code that significantly improves their lower bound. Their algorithm hinges on the use of total variation (TV) on indicator functions as a measurement of perimeter. Their discretized TV comes from \cite{Chambolle2016}, which we find converges poorly to perimeter. Instead, we implement a  discretization based on smoothed second order finite differences: \begin{equation}
    \frac{\partial}{\partial x} = 
    \frac{1}{6 \cdot dx}
    \begin{bmatrix}{}
    -1 & 0 & 1 \\
    -1 & 0 & 1 \\
    -1 & 0 & 1 
    \end{bmatrix}{},
    \hfill
    \frac{\partial}{\partial y} = 
    \frac{1}{6 \cdot dy}
    \begin{bmatrix}{}
    \;\;\;1 & \;\;\;1 & \;\;\;1 \\
    \;\;\;0 & \;\;\;0 & \;\;\;0 \\
    -1 & -1 & -1 
    \end{bmatrix}{}
\end{equation}{}
The total variation is then $\sum_{i,j=1}^n \left\|\frac{\partial u}{\partial x}, \frac{\partial u}{\partial y}\right\|_2 dx dy$. A comparison of the different discretizations at perimeter estimation are shown in Figure \ref{fig:tvdisc}. We also tested a smoothed third order finite difference gradient but find that the second order gradient achieves the closest perimeter estimate by far . We use this modified TV for all comparisons with \cite{DeFord2019TotalVI}. We sometimes obtain results where the lower bound intersects the upper bound. This is because even our modified TV does not exactly converge pointwise to continuous TV. Design of such a discretization is the subject of active research and is a challenge with TV-based algorithms more broadly. 

\begin{figure}
    \centering
    \includegraphics[width=.9\columnwidth]{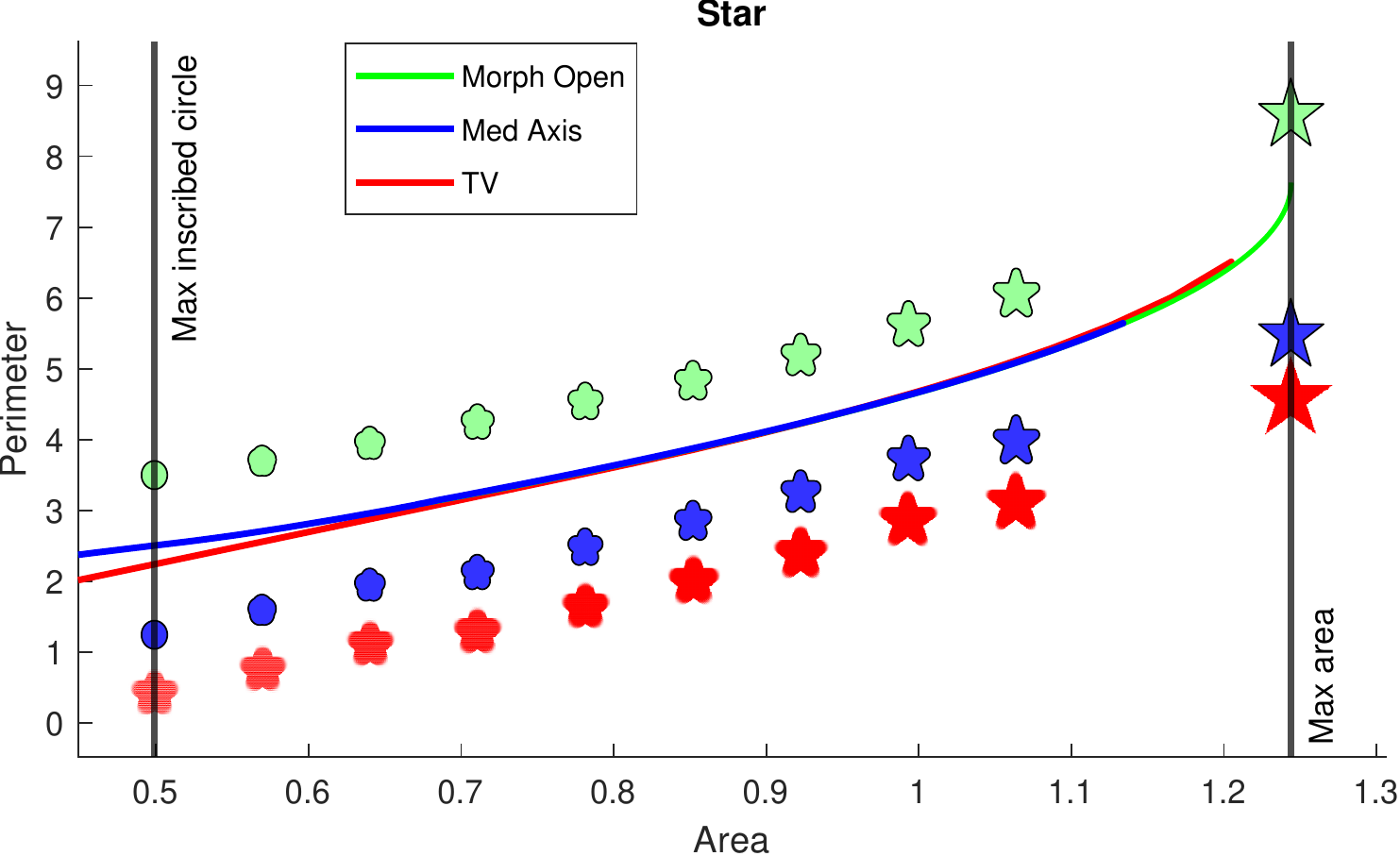}
    \includegraphics[width=.9\columnwidth]{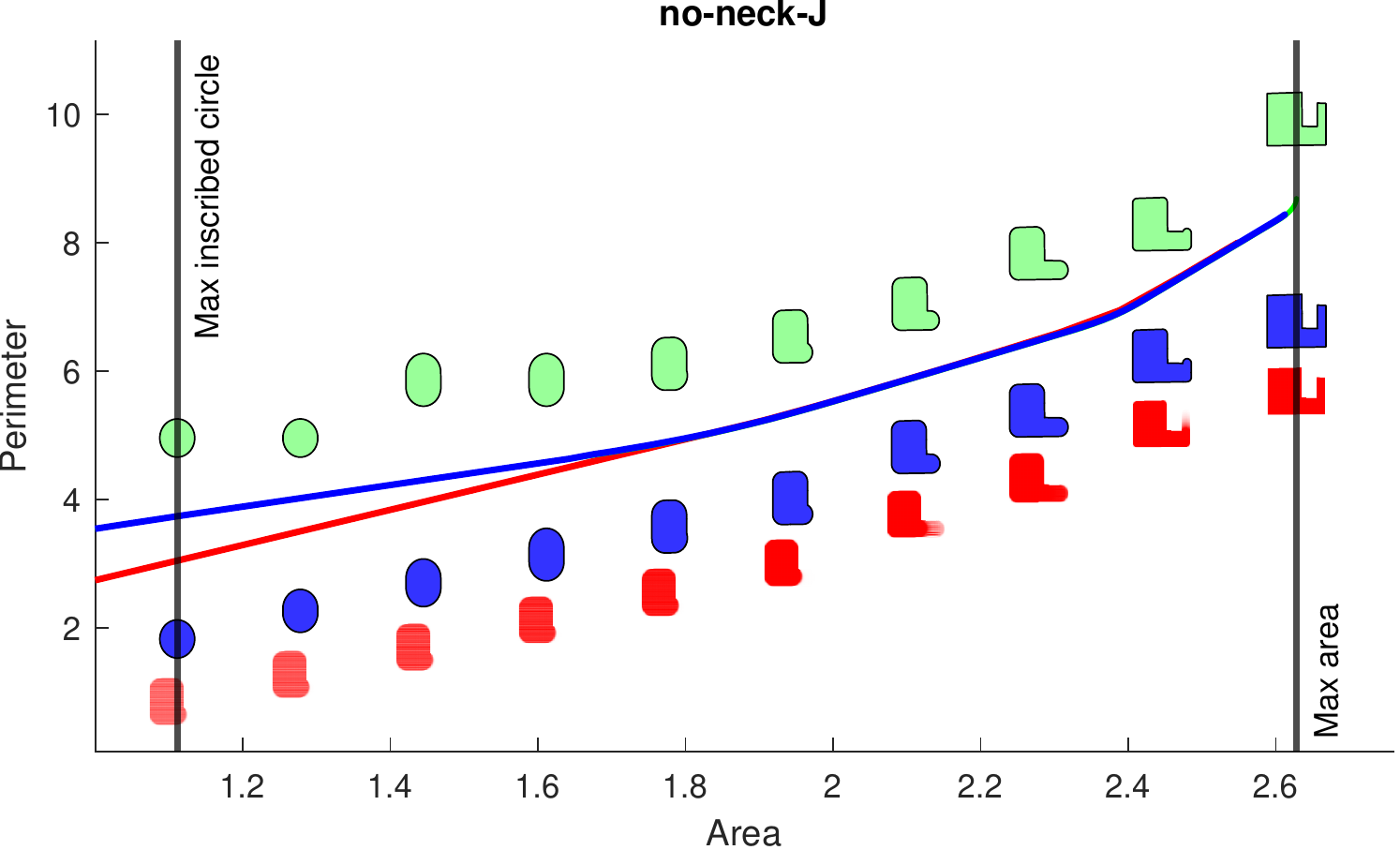}
    \caption{We compute three different isoperimetric profile bounds on domains with no neck: \textbf{Star}, \textbf{no-neck-J}. Both morphological opening and our medial axis algorithms produce tight upper bounds to the IP. TV roughly captures the convex lower envelope of the profile but slightly overshoots due to discretization error. }
    \label{fig:noneck}
\end{figure}{}

\begin{figure}
    \centering
    \includegraphics[width=.9\columnwidth]{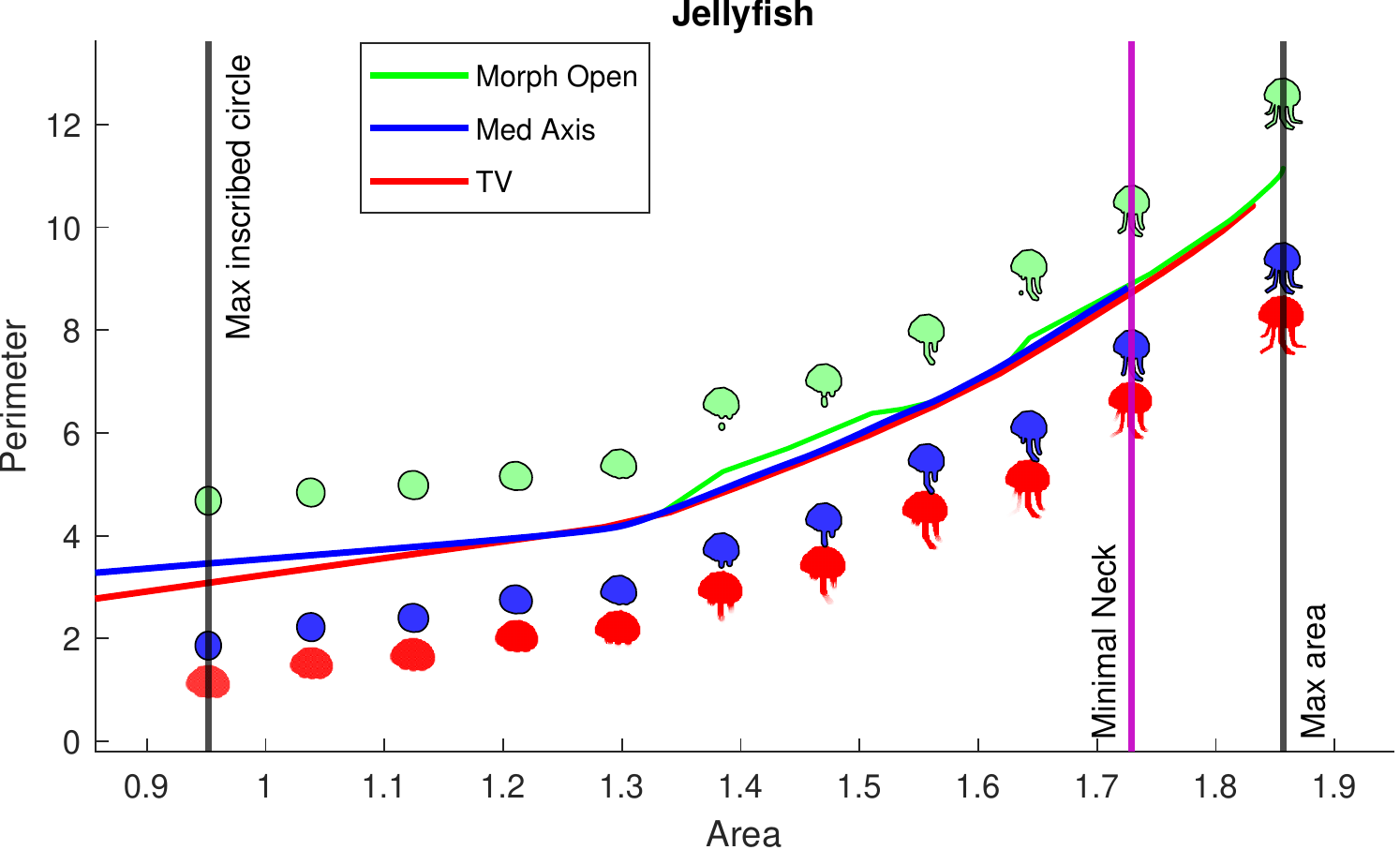}
    \includegraphics[width=.9\columnwidth]{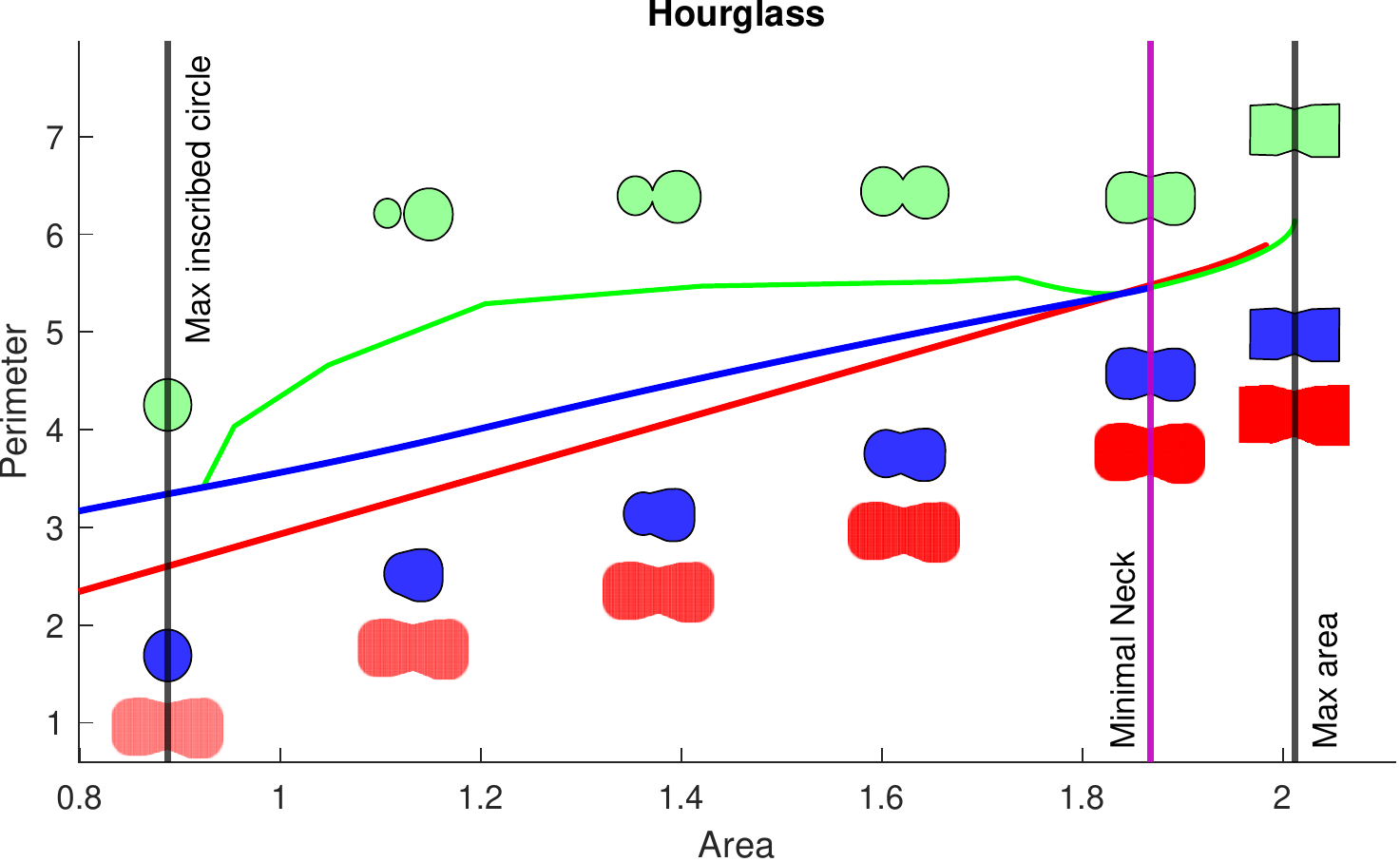}
    \includegraphics[width=.9\columnwidth]{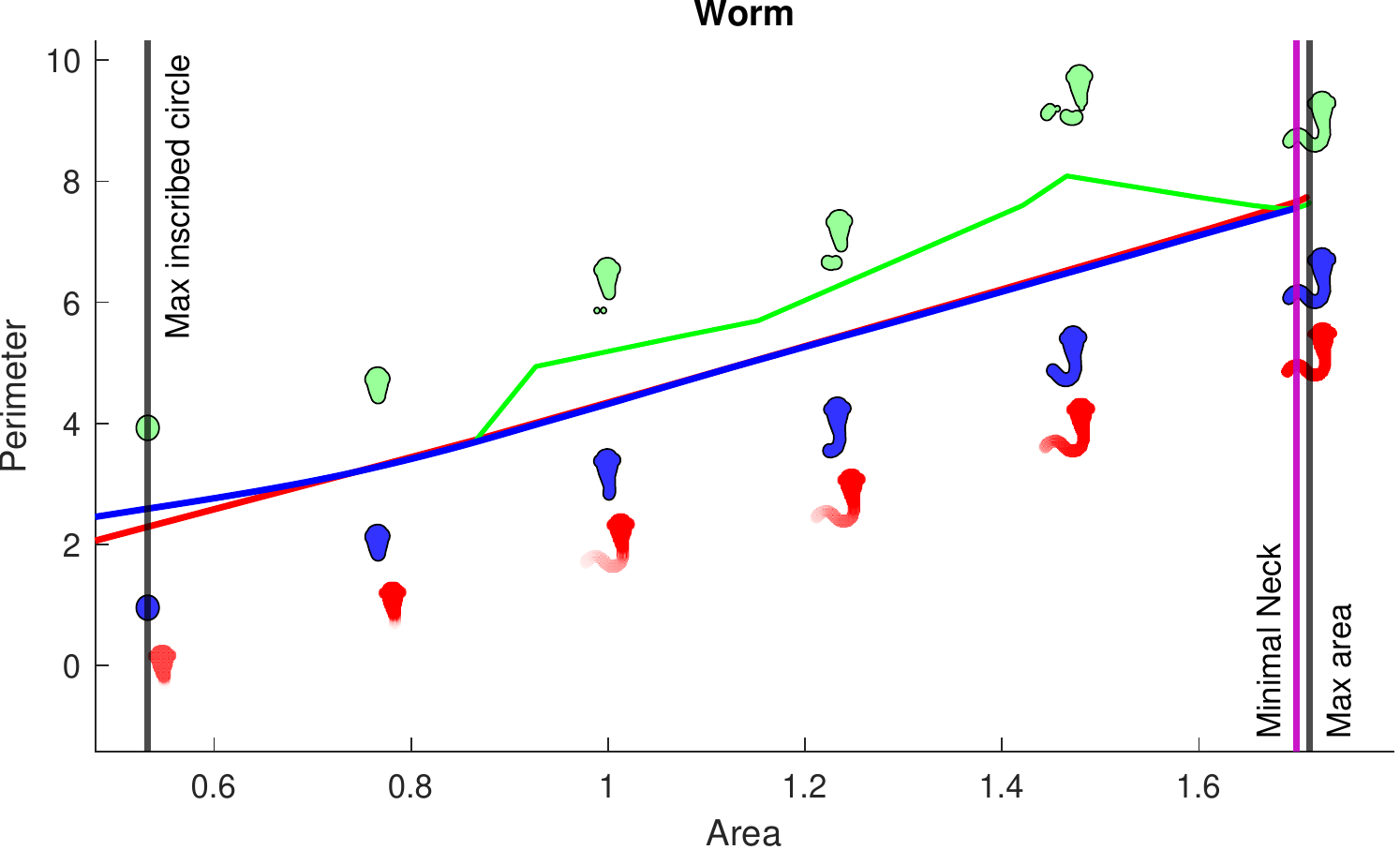}
    \caption{We compute three different isoperimetric profile bounds on domains with thick necks: \textbf{Jellyfish}, \textbf{Hourglass}, and \textbf{Worm}. 
    Morphological opening computes loose upper bounds, and TV computes loose lower bounds. While morphological opening can produce the exact IP on the small interval between ``Minimal Neck'' and ``Max area'', only our medial axis based algorithm computes the exact IP for the entire range of areas. }
    \label{fig:thicknecks}
\end{figure}{}

\subsection{No Neck Domains}
\label{sec:results:noneck}
On no neck domains both the morphological opening procedure and our medial axis traversal will produce the exact isoperimetric profile. This is depicted in Figure \ref{fig:noneck}. We test on a star domain \textbf{Star} as well as a non-star, no neck domain \textbf{no-neck-J}. \add{We choose }$r_l=.1$ \add{which} computes most of, but not the entire, profile\add{. This} \delete{which} is why our profile stops before the maximum area is reached. The total variation lower bound roughly follows the convex lower envelope of our profile, but overshoots in a few regions. As a lower bound, it leaves slack room, while our algorithm is exact.

\add{Note that the choice of $r_l=.1$ is specific to the domains we tested on. As $r_l$ roughly corresponds to the smallest length scale of a domain our IP bound will capture, a smaller domain requires a smaller $r_l$.}

\subsection{Thick Neck Domains}
\label{sec:results:thickneck}
On thick neck domains our algorithm computes the exact IP, while morphological opening computes a much looser upper bound, and total variation computes the convex lower envelope of the profile. This is demonstrated in Figure \ref{fig:thicknecks} on the \textbf{Jellyfish}, \textbf{Hourglass}, and \textbf{Worm} domains. Note that the ``Minimal Neck'' area, $\area(\Omega \circ r_n)$, is often very close to the maximum area. Prior methods would only be able to compute the exact profile in the limited region from $\area(\Omega \circ r_n)$ to $\area(\Omega)$, while our method computes the entire profile. Due to the prescence of necks, direct morphological opening produces much worse upper bounds than our method.

\subsection{General Domains}
\begin{figure*}
    \centering
    \includegraphics[width=.49\textwidth]{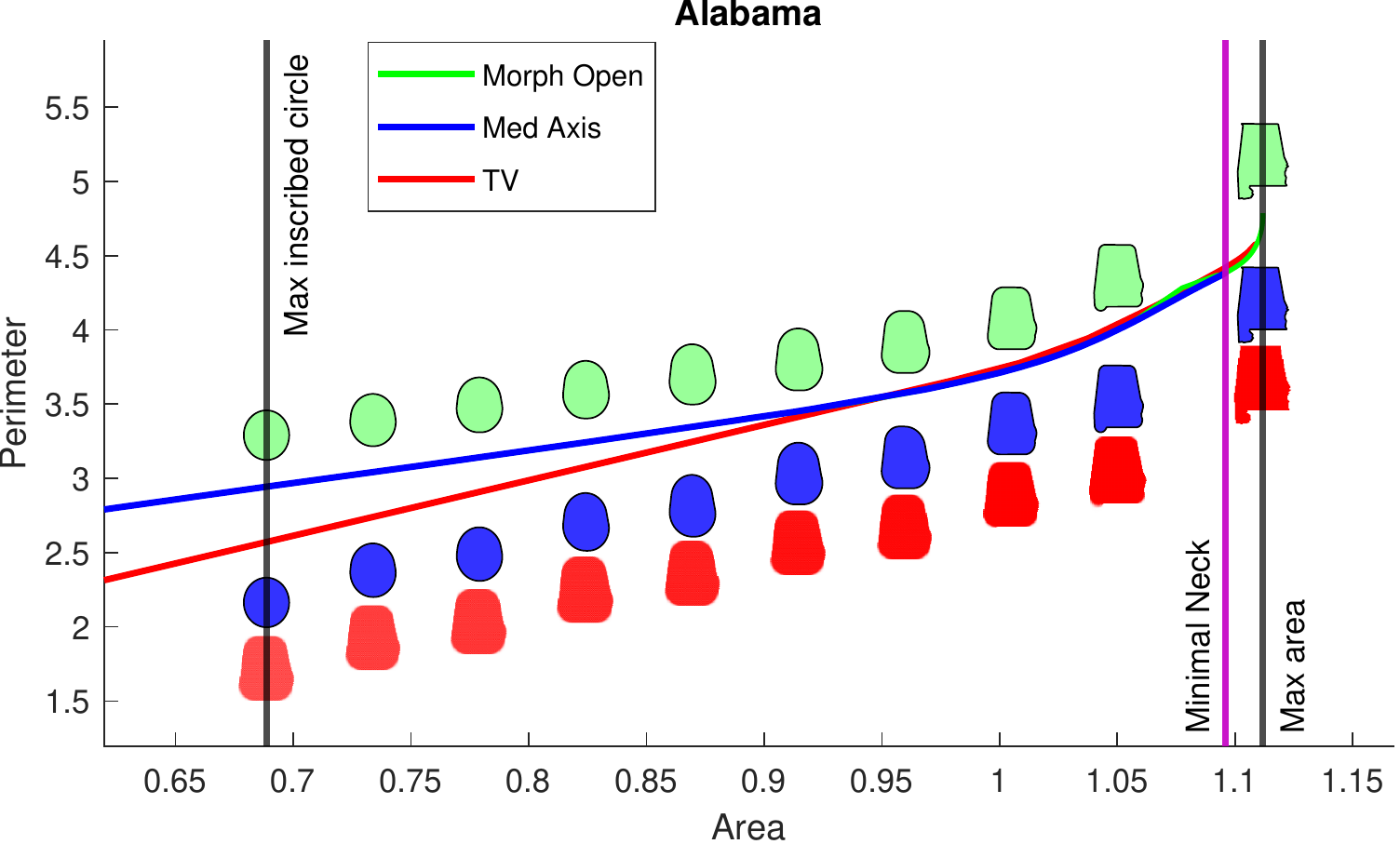}
    \hfill
    \includegraphics[width=.49\textwidth]{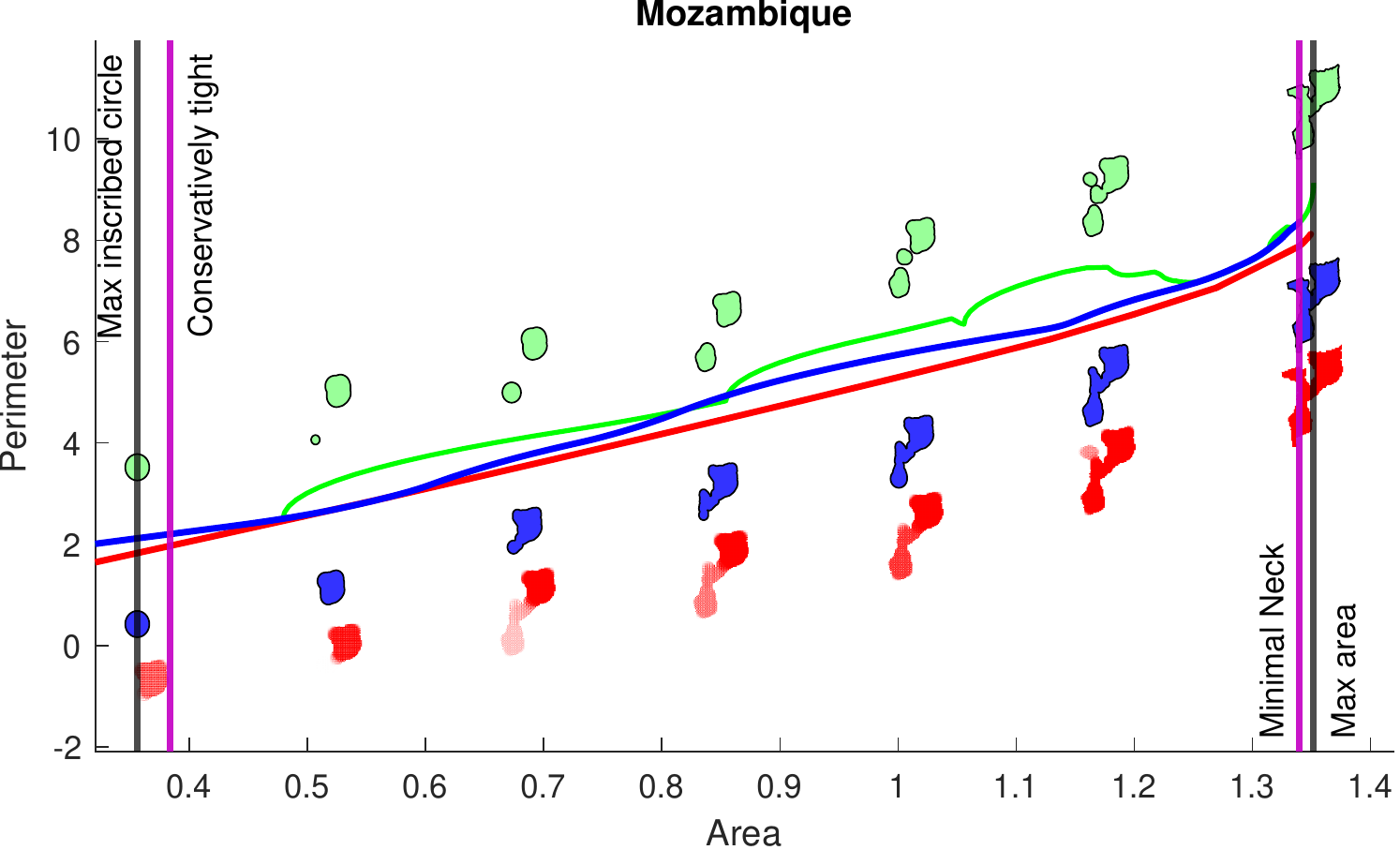}\\
    \includegraphics[width=.49\textwidth]{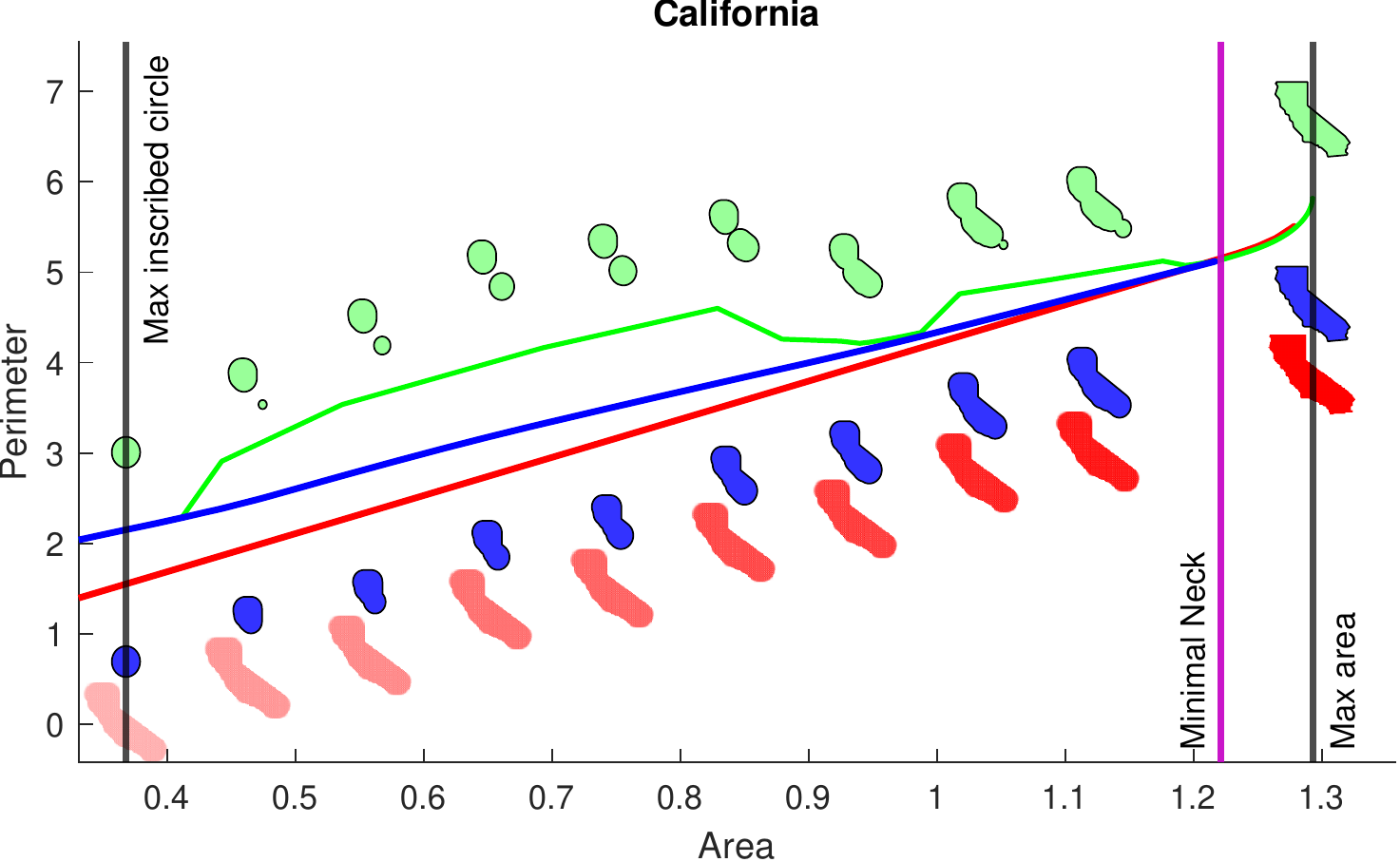}
    \hfill
    \includegraphics[width=.49\textwidth]{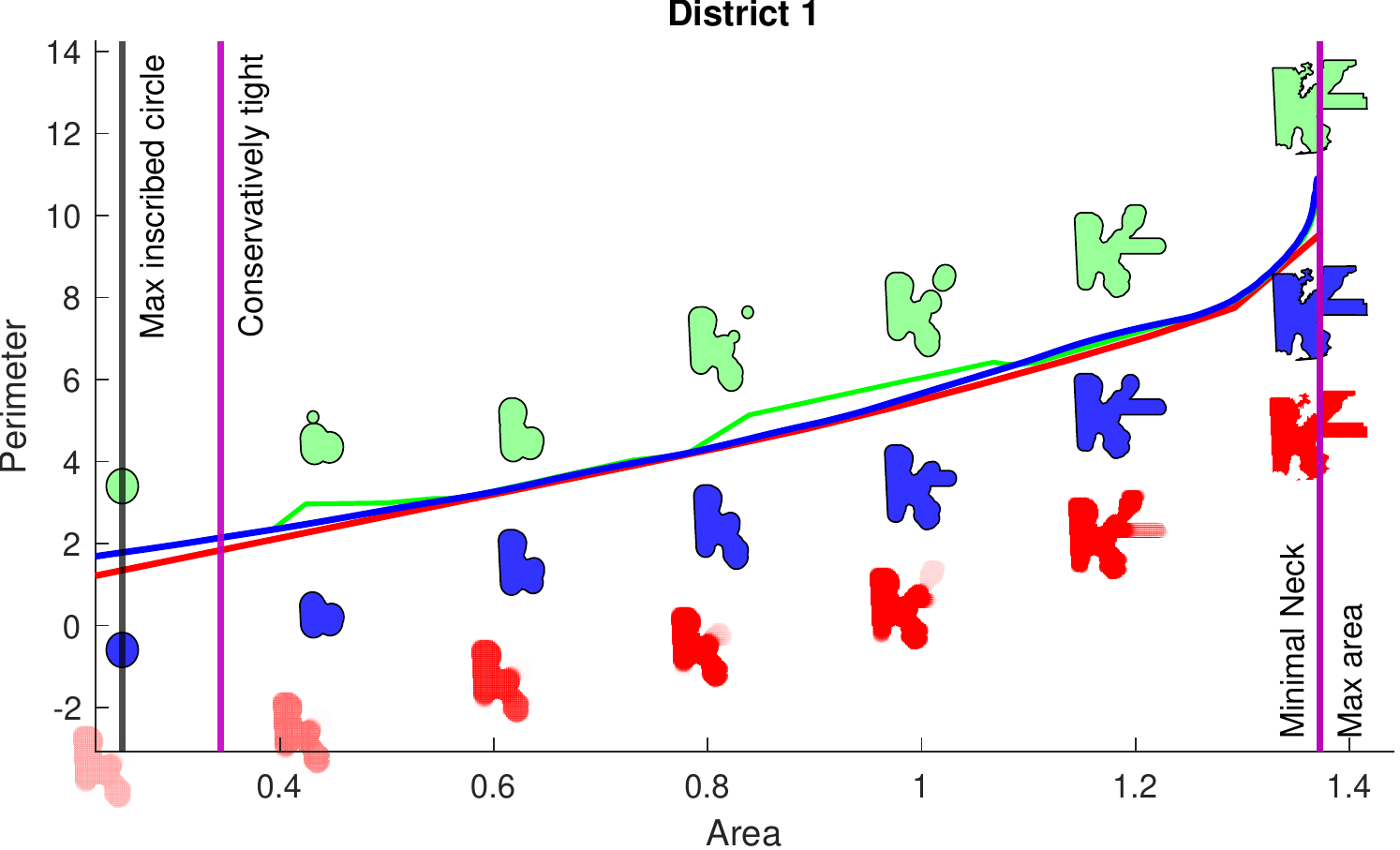}\\
    \includegraphics[width=.49\textwidth]{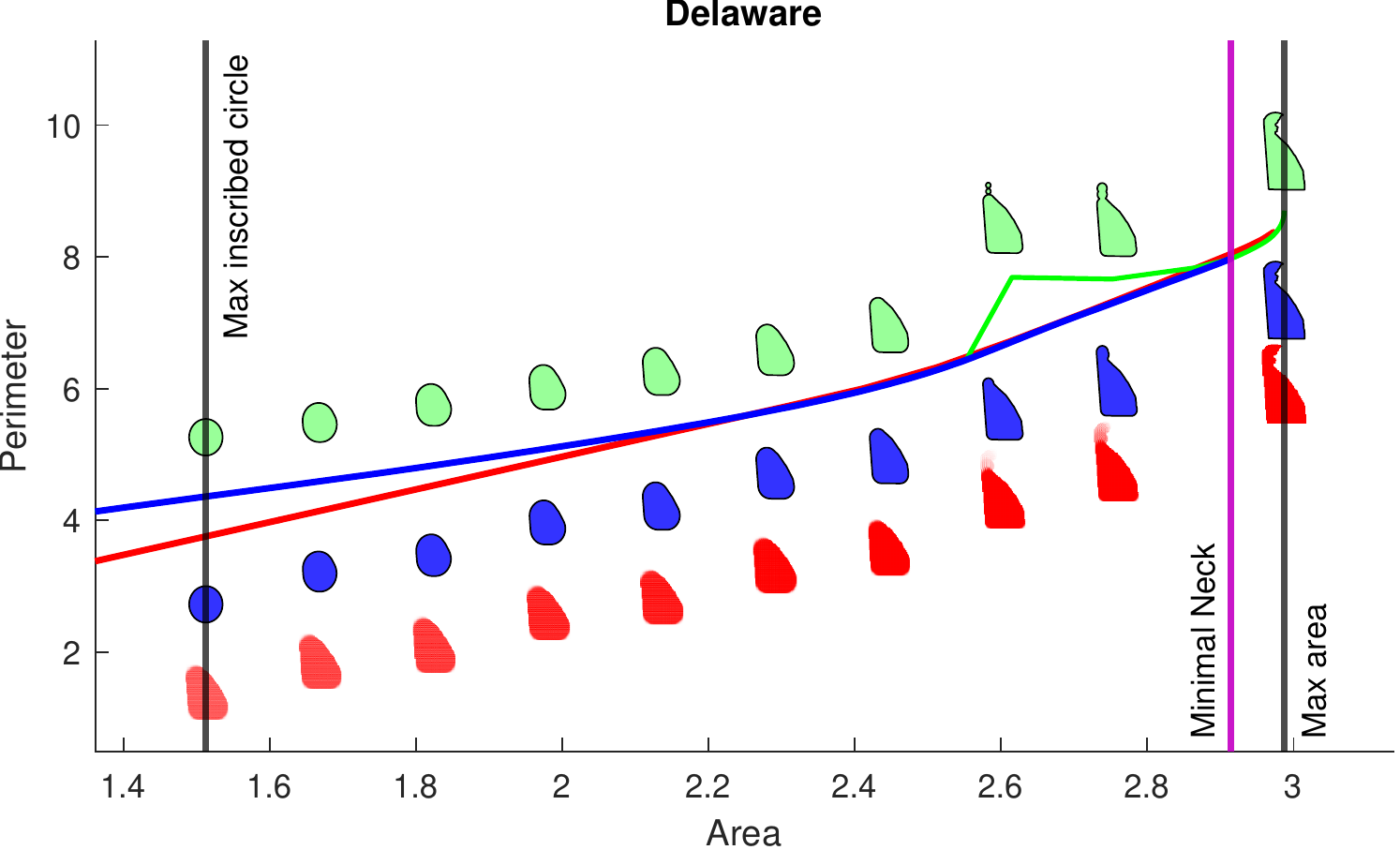}
    \hfill
    \includegraphics[width=.49\textwidth]{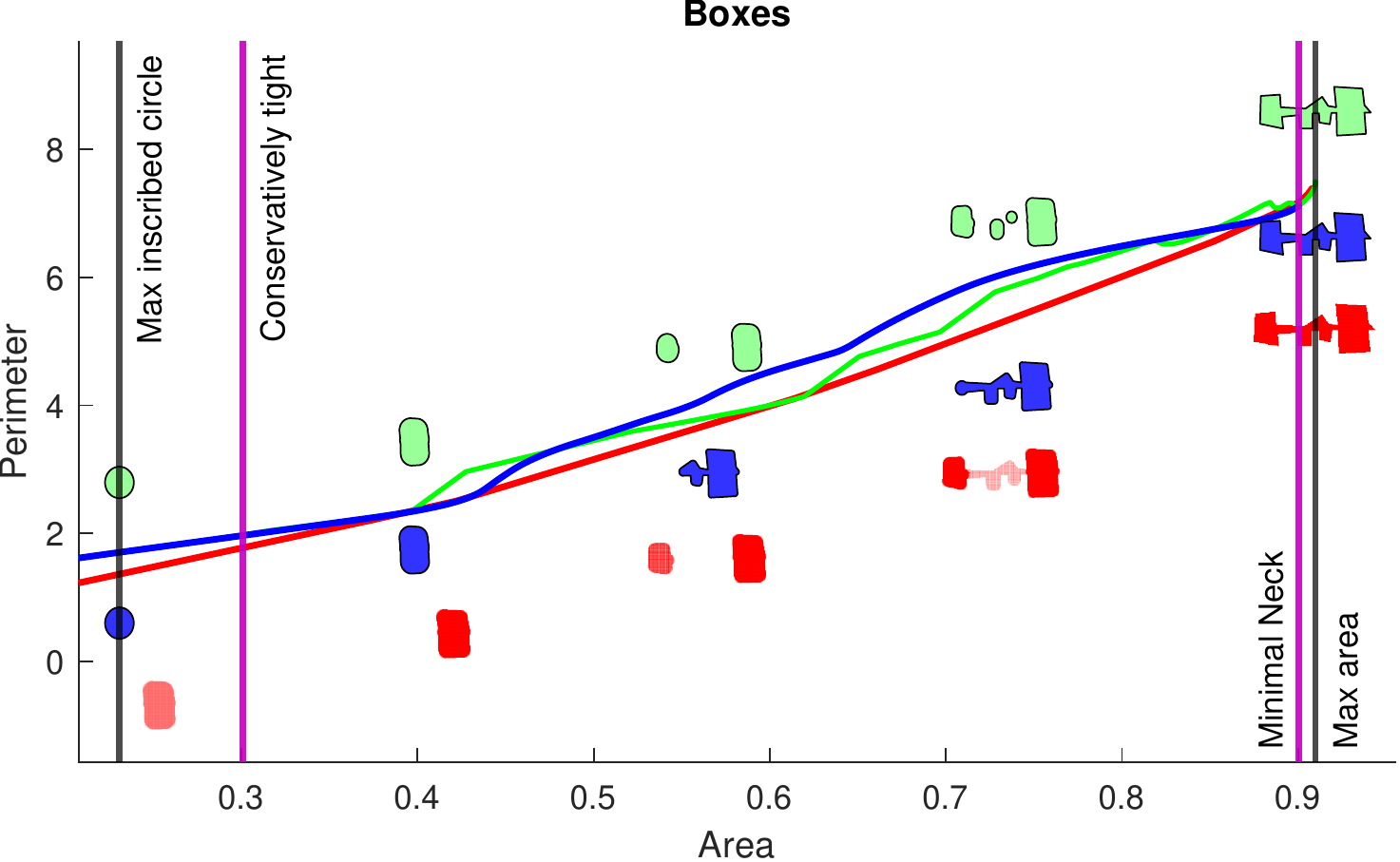}\\
    \caption{Isoperimetric profile bounds computed on a mix of sketch and map data. (Left column) \textbf{Alabama}, \textbf{California}, and \textbf{Delaware}. These U.S. states satisfy the thick neck condition so our upper bound is tight, while other bounds are significantly looser. (Right column) \textbf{Mozambique}, \textbf{District 1}, and \textbf{Boxes}. These domains do not satisfy the thick neck condition. Our bound still provides a tighter upper bound than opening on \textbf{Mozambique}, and \textbf{District 1}, but is looser on \textbf{Boxes}. This looseness is explained by the optimal inscribed shape being disconnected, something our algorithm can not achieve by construction. By Proposition \ref{prop:generaldomain}, our bound is tight to the left of ``Conservatively tight'' and to the right of ``Minimal Neck''. }
    \label{fig:generalIP}
\end{figure*}

We test on a set of geographic and hand drawn data. Maps of 
\textbf{Alabama}, \textbf{California}, \textbf{Delaware}, and \textbf{District 1} of Alabama were obtained from \cite{statefips}. 
A map of \textbf{Mozambique} was obtained from \cite{mozambique}. In addition, we create a drawing, \textbf{Boxes}, to test on. \textbf{Alabama}, \textbf{California}, and \textbf{Delaware} satisfied the thick neck condition while \textbf{District 1}, \textbf{Mozambique}, and \textbf{Boxes} do not. Our profiles for these domains are shown in Figure \ref{fig:generalIP}. Our bound computes the exact IP for all $t$ on \textbf{Alabama}, \textbf{California}, and \textbf{Delaware}. \cite{DeFord2019TotalVI} continues to provide convex lower envelopes with slack and an occasional intersection with the upper bound due to imperfect discretization. \add{This lower bound is particularly loose, on \textbf{Alabama}, and \textbf{California} revealing a significant gap for most of the profile.} Morphological opening produces upper bounds as well, but are loose compared to our medial axis based approach and has no theoretical guarantees except for a small region on the tail of the profile. 

On the \textbf{District 1}, \textbf{Mozambique}, and \textbf{Boxes} domains, we use Proposition \ref{prop:generaldomain} to compute ``Conservatively tight'' and ``Minimal Neck'' events that allow us to guarantee our bound is tight on a larger range of $t$ values. For both \textbf{District 1}, and \textbf{Mozambique}, our algorithm produces much tighter upper bounds than morphological opening. An exception occurs on \textbf{Boxes}, where the morphological opening produces a tighter bound. This is explained by the fact that the optimal $E_{\Omega}(t)$ is disconnected for much of the profile, while our algorithm by construction only produces connected polygons. For this reason it would be worthwhile to explore using disconnected medial axis reconstructions based on a criteria like the thick neck condition. Further exploration of disconnected \delete{partial}\add{limited} reconstructions is discussed in \S \ref{sec:discuss} and left to future work.

\vspace{-10px}
\subsection{Aggregate Profile}
\begin{figure}
    \centering
    \includegraphics[width=1\columnwidth]{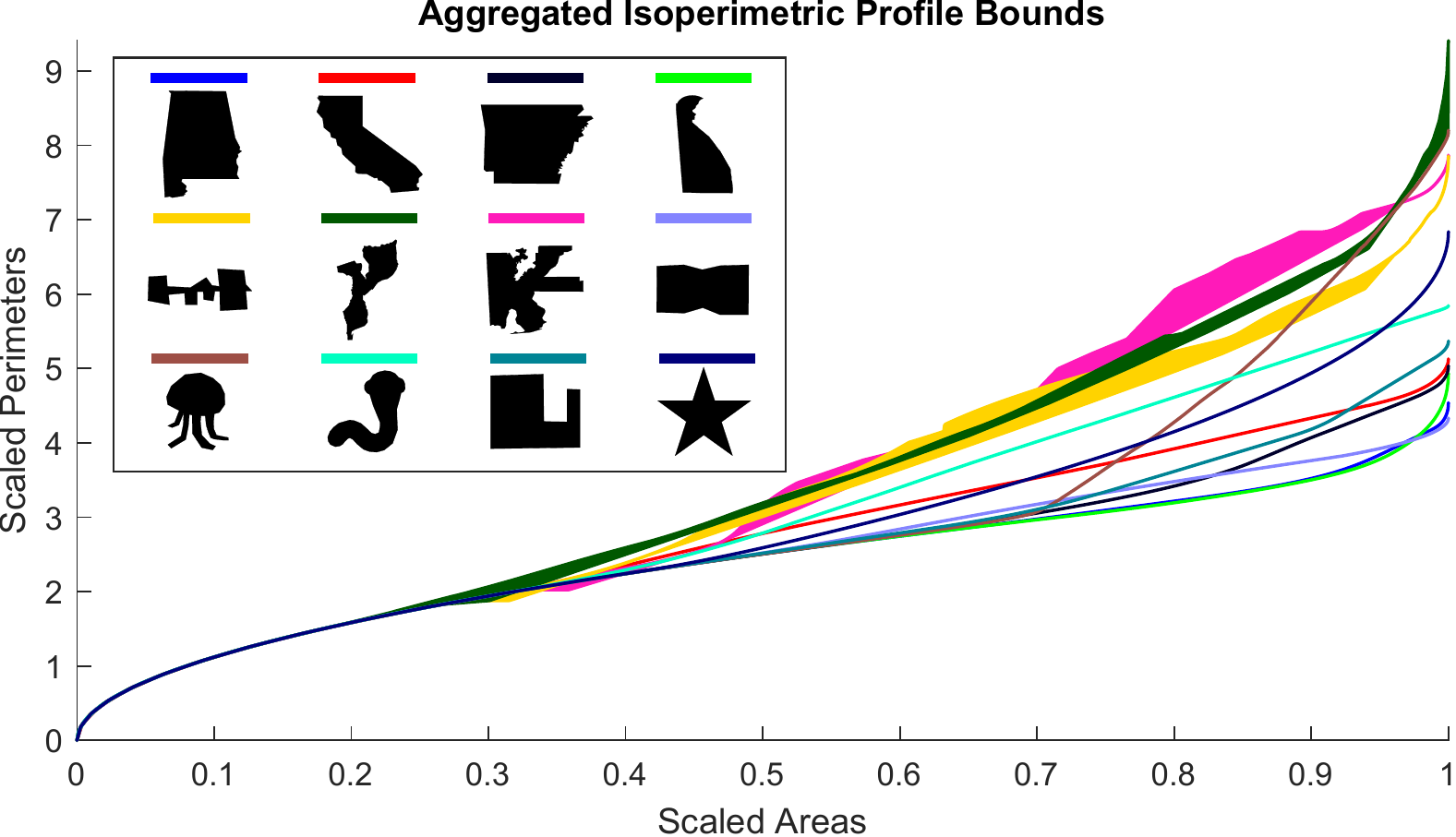}
    \caption{We combine bounds and guarantees of this paper and those of \cite{DeFord2019TotalVI, giorgio2019} to produce a tight regime inside which the IP must lie. Domains are scaled to have area $1$ for plotting purposes. We find that this produces a relatively narrow range of uncertainty that illuminates the rough shape of the IP. This plot also allows us to see the rich space of multi-scale phenomena that 2D domains can have. In particular, the (brown) \textbf{Jellyfish}'s profile starts shallow before sharply veering upwards, reflecting the composition of the \textbf{Jellyfish} by a big round cap and many small tentacles.}
    \label{fig:aggIP}
\end{figure}

By combining the bounds and guarantees provided in this paper and those of \cite{DeFord2019TotalVI, giorgio2019}, we compute the envelope inside which the isoperimetric profile must lie. This is visualized in Figure \ref{fig:aggIP}. We find that this produces a relatively narrow region of uncertainty where the IP is unknown. This visualization also makes clear the rich space of multi scale behaviors \add{that} different domains can have. 

For example, consider the \textbf{Jellyfish} (brown). It consists of a large cap region which has mostly low resolution features, followed by several tendrils that require high resolution data to capture. This is reflected in its isoperimetric profile where it starts shallow for most of the area, before sharply veering up to have one of the largest perimeters in this figure.  
\add{\textbf{Delaware} (bright green) demonstrates similar behavior to a lesser extent. It starts with a shallow profile indicating its coarse lower portion, before its profile curves upwards to capture the high resolution details of its upper quarter. \textbf{Delaware} requires less high resolution data to capture than the \textbf{Jellyfish}, however, as indicated by its profile sitting below that of the \textbf{Jellyfish}.}
Our results with the isoperimetric profile make it possible to computationally and numerically \add{compare these} \delete{quantify a} domains' multi-scale properties.
{}

{}

\vspace{-10px}
\subsection{Runtime}
A list of runtimes for our experiments is provided in Table \ref{table:timings}. The runtime of our algorithm depends most strongly on number of vertices $|V|$, and loosely on other parameters $d_c$, $r_l$, and $r_n$. For morphological opening, the runtime depends primarily on the number of samples of the IP that are computed. For all times listed, we sample 100 values of the IP. For TV, runtime depends on the number of samples of the IP that are computed as well as the pixel resolution used to rasterize the polygon. We compute 20 samples of the TV IP and use 100 vertical pixels for the rasterization. 

{}

\begin{table}[]
\scalebox{.9}{
\begin{tabular}{| c | c | c | c | c | c |}
\hline
Number of & Mesh     & MA  & Open  & TV  &   $r_n$     \\ 
Vertices  & name     & time (s) & time (s) & time (s) &             \\ \hline
9         & no-neck-J        & 29.99   &   6.185    & 26.6 &   $\infty$  \\ \hline
11        & Star     & 11.489   & 1.44   &  13.9  &   $\infty$  \\ \hline
14        & Candy     & 18.694   & 7.72   &  115.9  &   0.444   \\ \hline
43        &   Delaware   & 303.52   & 7.69    & 24.4  &   $\infty$  \\ \hline
46        & Jellyfish    & 156.81   & 5.70    & 165.3  &   0.069  \\ \hline
146       & \small{Mozambique}  & 1251.1   & 18.50    & 10.6  &   0.040  \\ \hline
\end{tabular}}
\caption{Runtimes of our method, and comparison methods in seconds. Since our method is Lagrangian using floating precision geometry, its runtime is correlated with number of vertices. 
Total variation uses a pixelization of the domain, and we choose 100 pixels vertically for all experiments listed. Morphological opening is also \delete{l}\add{L}agrangian and shows slight correlation with number of vertices. Our runtime is comparable to the runtime of TV on moderately sized polygons.}
\label{table:timings}
\end{table}

\vspace{-10px}
\section{Discussion and Conclusion}
\label{sec:discuss}
This work enables computation of a significant portion of the isoperimetric profile, a far more robust measurement of geometric compactness than the popularly used isoperimetric ratio. We stress that while more challenging to compute, a multi-scale approach to isoperimetry is necessary to perform reliable shape analysis. By providing an upper bound counterpart to \cite{DeFord2019TotalVI} we hope to lower the barrier between users and multi-scale measurement of geometric compactness.

\add{ This complementary bound and the relatively small gap between them in many of our examples as in  Figure \ref{fig:aggIP} provides significantly increased confidence that the IP can be used in practice for analyzing potential districting plans. While the examples in  \cite{DeFord2019TotalVI} demonstrated that the IP could reveal multiscale structure in individual districts and highlight distinct types of undesirable behavior in district shapes, the fact that the TV relaxation only recovers the lower convex envelope means there was no guarantee that the computed function was representative of the actual IP. 

The quite tight IP bounds of \textbf{District 1} in Figure \ref{fig:generalIP} give an example of several features common among real districting plans, including the small size of the maximum inscribed circle and rapid acceleration of the IP for large areas as the fine details are filled in. In addition to the problems of coastline and resolution discussed in the introduction, districting plans are usually constructed out of discrete units like census blocks and hence the map drawer may have little control over the finest level of details, which are most heavily penalized by scores like Polsby-Popper that report only a single number to summarize the compactness of the plan. In between the extremal regions, the profile recovers the relative size of necks and tendrils, which are in some cases more likely markers of gerrymandering, for example by packing together urban centers, 
than rough boundary elements. 
 }
 
 
While we have made several advancements on this problem, there remain many avenues in which to approach further improvement.  On the theoretical side, we conjecture that $E_{\Omega}(t)$ can always be constructed from a union of balls centered on the medial axis. None of our many experiments have demonstrated otherwise. Proof of this conjecture would decisively reduce the complexity of this problem from 2D to just 1D. 

Empirically, our medial axis based algorithm suffers when the optimal solution has disconnected $E_{\Omega}(t)$. Many heuristics are available for constructing disconnected medial axis reconstructions and merit further exploration. More recent work in computer graphics have used semidefinite relaxations of quadratic problems with great success. This could be formulated into a method of selecting a subset of all balls centered on the medial axis for the reconstruction. A more brute force approach could be to compute the next largest inscribed circle at each iteration and compare that with connected traversal of the medial axis. Lastly, our definition of necks partitions $\Omega$ into a union of no neck subregions. This suggests a coarser scale optimization where the atomic unit is a no neck subregion rather than just a ball.

On the lower bound side, \cite{DeFord2019TotalVI} produces a convex relaxation of the problem where mass density can take values between 0 and 1. A simple way to tighten this lower bound is just to run a subsequent nonlinear optimization initialized at the relaxed solution. This would be reminiscent of techniques used in topology optimization for mass constrained mechanical part design.

Our medial axis based approach can be na\"ively extended to 3D, where it also produces upper bounds. Unfortunately in the 3D setting $\partial E_{\Omega}(t)$ is not necessarily a spherical shell but rather a minimal surface. It would be interesting to measure how big the gap between upper and lower bounds are in the 3D setting.

\add{The isoperimetric profile can be used to distinguish between shapes that are intrinsically identical. For example, \cite[Figure 15]{gordon1996you} define different shapes $D_1$ and $D_2$ that have identical spectra. However, their isoperimetric profiles are guaranteed to be distinct since $\inr(D_1)\neq\inr(D_2)$. This simple demonstration incentivizes further exploration of the isoperimetric profile as a shape descriptor. 
}


By relating three classic geometry processing concepts, the medial axis, morphological opening, and isoperimetry, we achieve significant progress towards computation of the isoperimetric profile. 

\section*{Acknowledgements}
The authors thank Edward Chien, Giorgio Saracco, and Hugo Lavenant for many valuable discussions. 
Paul Zhang acknowledges the generous support of the Department of Energy Computer Science Graduate Fellowship and the Mathworks Fellowship. 
Daryl DeFord and Justin Solomon acknowledge the generous support of the Prof.\ Amar G.\ Bose Research Grant. 
Justin Solomon and the MIT Geometric Data Processing group acknowledge the generous support of Army Research Office grants W911NF1710068 and W911NF2010168, of Air Force Office of Scientific Research award FA9550-19-1-031, of National Science Foundation grant IIS-1838071, from the MIT--IBM Watson AI Laboratory, from the Toyota--CSAIL Joint Research Center, from a gift from Adobe Systems, and from the Skoltech--MIT Next Generation Program.

\appendix
\vspace{-10px}
\section{Proof of Proposition \ref{prop:marecon}, $\Omega_{g_{\rho}}=\Omega \circ B_{\rho}$}
\label{proof:medaxrecon}
\begin{proof}
Recall that by definition $g_{\rho} = \{x\in\MA(\Omega): r(x) \geq \rho\}$ is the set of medial axis points whose radii are more than $\rho$. An alternative description for $\Omega \circ B_{\rho}$ is the union of all circles of radius $\rho$ that fit in $\Omega$. 
It suffices to show that $\Omega_{g_{\rho}}$ is also the union of all circles of radius $\rho$ that fit in $\Omega$. 

Given a point $x\in\Omega_{g_{\rho}}$, it must be contained in a ball of radius larger than or equal to $\rho$ in $\Omega$ because $\Omega_{g_{\rho}}$ is the union of such circles. Such a circle must be in $\Omega \circ B_{\rho}$ and so $x$ must be as well.

Start with a point $x\notin\Omega_{g_{\rho}}$. Consider the biggest circle in $\Omega$ that contains $x$. If the radius of this circle is larger than or equal to $\rho$, then this circle can be enlarged until it lies tangent to multiple boundary vertices while still containing $x$. The center of the enlarged circle would be on the medial axis of radius more than $\rho$, which contradicts $x\notin\Omega_{g_{\rho}}$. Therefore there is no circle of radius larger than or equal to $\rho$ that encloses $x$ i.e. $x\notin\Omega\circ B_{\rho}$.
\end{proof}{}

\vspace{-10px}
\section{Proof of Proposition \ref{prop:neckequality}: Neck Equality}
\label{proof:neckneck}
\begin{proof}
We start with the point $x_0\in\MA(\Omega)$ of minimal neck radius. We can assume $x_0$ satisfies condition $2$ in Definition \ref{def:neck2} because condition $1$ can be reframed into condition $2$ by adding an imaginary medial axis junction node at $x_0$. Label the two outgoing medial axis edge segments on which change in $r(x)$ is positive by $e_1$, and $e_2$. 

First we show that $\Omega$ has a neck of radius $r(x_0)$ by Definition \ref{def:neck}. This is constructed by taking $x_1,\;x_2$ on $e_1,\;e_2$ respectively an $\epsilon$ distance away from $x_0$. By Definition \ref{def:neck2}, $r(x_1)>r(x_0)$, and $r(x_2)>r(x_0)$. Since $\Omega$ is a Jordan domain, the path from $x_1$ to $x_2$ that stays furthest from $\partial\Omega$ must pass through the choke point $x_0$. Even then, at the choke point, a ball of radius $r(x_1)$ or $r(x_2)$ will not fit since $r(x_0)$ is smaller. Therefore, $\Omega$ contains a neck of radius $r(x_0)$

Next we show that $\Omega$ does not have a neck of radius $\rho < r(x_0)$. Given any $x_1, x_2$ and balls of radius $\rho<r(x_0)$ around them, we can construct $\gamma$ satisfying Definition \ref{def:neck} by the following. Split $\gamma$ into three parts. Let the first part be $\gamma_1$ the shortest path contained in $\Omega$ from $x_1$ to the medial axis. Denote the point where $\gamma_1$ contacts the medial axis by $x_1^m$. Let the third part be the reverse of the shortest path $\gamma_3$ contained in $\Omega$ from $x_2$ to the medial axis. Denote the point where $\gamma_3$ contacts the medial axis by $x_2^m$. Finally let the second part $\gamma_2$ be the path from $x_1^m$ to $x_2^m$ via the medial axis. Since $\gamma$ connects the two balls, $\Omega$ does not contain a neck of radius $\rho$.
\end{proof}{}

\vspace{-10px}
\section{Proof of Proposition \ref{prop:generaldomain}}
\label{proof:earlyT}
\begin{proof}
Starting from $\inB(\Omega)$, we can increase $E_{\Omega}(t)$ continuously until the influence of the first neck is reached. This corresponds to the first time $g$ absorbs a point on the medial axis of radius $r_m$. At this event, $\Omega_g = \Gamma_{r_m}\oplus B_{r_m}$. Before then, $E_{\Omega}(t)$ is expanding in a neckless domain and so, excluding the possibility of generating a new disconnected region, greedy traversal is optimal.

Taking into account the possibility of generating a new region, the earliest moment could happen is if the radius of curvature of $\partial E_{\Omega}(t)$ is equal to $\frac{\inr(\Omega)}{2}$. Therefore, to be conservative, let $r_t=\max(r_m,\frac{\inr(\Omega)}{2})$. Until $t = \area(\Gamma_{r_t}\oplus B_{r_t})$ it will be optimal to proceed as if in a no neck domain.
\end{proof}{}

\vspace{-10px}
\section{Proof of Proposition \ref{prop:thickneck}: Tight upper bound for thick neck domains assuming $E_{\Omega}(t)$ grows only continuously excluding disconnected circles}
\label{proof:bigneck}
\begin{proof}
We consider ways in which $E_{\Omega}(t)$ can increase. 

A new disconnected component could be generated anywhere in $\Omega - E_{\Omega}(t)$. In such a case, by the isoperimetric inequality, the optimal shape of the new component is a circle of maximal radius. For an optimistic measurement of how preferable this option is we replace $\Omega - E_{\Omega}(t)$ by $\Omega - \inB(\Omega)$. The radius of the next largest circle that could be inscribed in $\Omega - \inB(\Omega)$ is by definition $\inr_2(\Omega)$. The corresponding change in the isoperimetric profile is then 
$\frac{\length(\partial( \inB_2(\Omega)))}{\area(\inB_2(\Omega)))} = \frac{2}{\inr_2(\Omega)}$.

If no disconnected components are generated, a continuous growth in $E_{\Omega}(t)$ corresponds to increasing $g$ along the medial axis. As derived in equation $\eqref{eq:differentialscore}$, the slope this induces is $\frac{1}{r(x)}$. 

As long as $\frac{1}{r(x)}\leq\frac{2}{\inr_2(\Omega)}$, the IP will not prefer to generate a new disconnected region. While $t < \area(\Omega\circ B_{r_n})$, the largest value of $\frac{1}{r(x)}$ will occur at $\frac{1}{r_n(x)}$. Therefore if $r_n(x) \geq \frac{\inr_2(\Omega)}{2}$, then $E_{\Omega}(t)$ will always prefer traversing the medial axis. 

This proposition excludes cases where $E_{\Omega}(t)$ could transform discontinuously without increasing number of components and we conjecture that such a change would not be optimal. 
\end{proof}

\bibliographystyle{eg-alpha-doi} 
\bibliography{bibtex}


\end{document}